\documentclass[conference]{IEEEtran}

\usepackage[tight,footnotesize]{subfigure}
\usepackage{graphicx}
\usepackage{url}
\usepackage[ruled,noend]{algorithm2e}

\SetAlFnt{\small}
\SetAlCapFnt{\small}
\SetAlCapNameFnt{\small}
\SetAlCapHSkip{0pt}
\IncMargin{-\parindent}

\usepackage{caption}
\usepackage{array}
\usepackage{amsmath}
\usepackage{amsxtra}
\usepackage{amsfonts}
\usepackage{hyperref}
\usepackage{booktabs}

\IEEEoverridecommandlockouts

\begin{document}

\title{Texture to the Rescue: Practical Paper Fingerprinting based on Texture Patterns\thanks{This manuscript has been accepted for publication in the ACM Transactions on Privacy and Security (TOPS, formerly TISSEC) in 2017.}}



\author{\IEEEauthorblockN{Ehsan Toreini}
\IEEEauthorblockA{School of Computing Science\\
Newcastle University\\
United Kingdom\\
ehsan.toreini@ncl.ac.uk}
\and
\IEEEauthorblockN{Siamak F. Shahandashti}
\IEEEauthorblockA{Department of Computer Science\\
University of York\\
United Kingdom\\
siamak.shahandashti@york.ac.uk
}
\and
\IEEEauthorblockN{Feng Hao}
\IEEEauthorblockA{School of Computing Science\\
Newcastle University\\
United Kingdom\\
feng.hao@ncl.ac.uk
}
}

\maketitle


\begin{abstract}

In this paper, we propose a novel paper fingerprinting technique based on analyzing the translucent patterns revealed when a light source shines through the paper. These patterns represent the inherent texture of paper, formed by the random interleaving of wooden particles during the manufacturing process. We show these patterns can be easily captured by a commodity camera and condensed into to a compact 2048-bit fingerprint code. Prominent works in this area (Nature 2005, IEEE S\&P 2009, CCS 2011) have all focused on fingerprinting paper based on the paper ``surface''. We are motivated by the observation that capturing the surface alone misses important distinctive features such as the non-even thickness, the random distribution of impurities, and different materials in the paper with varying opacities. Through experiments, we demonstrate that the embedded paper texture provides a more reliable source for fingerprinting than features on the surface. Based on the collected datasets, we achieve 0\% false rejection and 0\% false acceptance rates. We further report that our extracted fingerprints contain 807 degrees-of-freedom (DoF), which is much higher than the 249 DoF with iris codes (that have the same size of 2048 bits). 
The high amount of DoF for texture-based fingerprints makes our method extremely scalable for recognition among very large databases; it also allows secure usage of the extracted fingerprint in privacy-preserving authentication schemes based on error correction techniques. 



\end{abstract}


\IEEEpeerreviewmaketitle

\section{Introduction}
\label{introduction}
{\bf Secure paper documents.}
Designing secure documents that provide high levels of security against physical forgery is a long-standing problem. Even in today's digital age, this problem remains important as physical paper is still prevalently used in our daily lives as a means to prove data authenticity, for example, in receipts, contracts, certificates, and passports. A recent trend in this area (e.g., in e-passports) is to embed electronics such as RFID chips within the physical document in question \cite{SecurityEPassports}. However, the security of such solutions depends on the tamper-resistance of the chip which must securely store a long-term secret \cite{tracibilityEPassports}. This tamper-resistance requirement can significantly increase the cost of production. In view of the importance of ensuring the authenticity of paper documents, researchers have been exploring applying digital technologies to prevent counterfeiting. One promising method is based on measuring the unique physical properties of paper that are impossible to clone.


\textbf{Paper fingerprinting.}
Manufacturing a paper sheet is a complex process and each paper sheet is a unique product from that process. Typically, wooden particles are used as the base, and multiple substances are subsequently applied to stick these particles together to stabilize their placement and shape a thin, usually white, steady surface which we call paper. 

In an article published in \emph{Nature} in 2005, Buchanan et al.\ observed that the surface of a paper sheet is imperfect -- it contains random non-evenness as a natural outcome of the paper manufacturing process~\cite{paperPUF}. They propose to utilize the surface imperfections to uniquely identify the paper. Their method is to use a focused laser beam to scan a pre-designated region on the paper sheet from four different angles, and continuously record the intensity of the reflected laser. The recordings then constitute a unique digital representation of the paper, called ``paper fingerprint''. 
Therefore, Buchanan et al.'s method~\cite{paperPUF} is the basis of a number of follow-up works, notably~\cite{sampleCorrelationLaser,laseRecognition}. 

Clarkson et al.~(IEEE S\&P, 2009) subsequently showed that a commodity scanner could be used to effectively extract paper fingerprints based on the same surface imperfections~\cite{clarksonfingerprint}. Their method is to scan the paper surface from four different angles and then construct a 3-D model. Then the 3-D model is condensed into a concise feature vector, which forms the paper fingerprint. 

Later, Sharma et al.~(CCS, 2011) proposed another approach named PaperSpeckle, which uses a microscope with a built-in LED as the light source to extract the paper speckle patterns at the microscopic level (1--2 microns) \cite{paperspeckle}. The underlying idea in PaperSpeckle is based on the concept of speckles: i.e., when light falls on a paper sheet, the scattered light forms randomly mixed bright and dark regions, which can then be captured by a microscope. The captured image can be further processed to produce a compact binary fingerprint.

\textbf{Our idea.} 
So far, prominent works in this area have primarily focused on the imperfections of the paper surface. 
In contrast, our work is inspired by the observation that the wooden particles constituting the building blocks of a paper sheet scatter over the paper quite irregularly. 
We hypothesize that this irregular placement of wooden particles provides a unique pattern, which can be extracted and used as a paper fingerprint. We call the unique pattern caused by the random interleaving of wooden particles the \emph{texture} of paper. 



\begin{figure}
	\centering
	\subfigure[Paper Surface]
	{
		\includegraphics[scale=0.2]{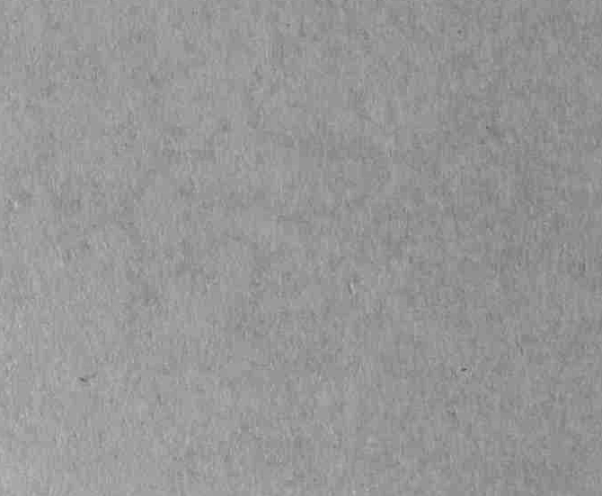}
	}
	\subfigure[Paper Texture]
	{
		\includegraphics[scale=0.2]{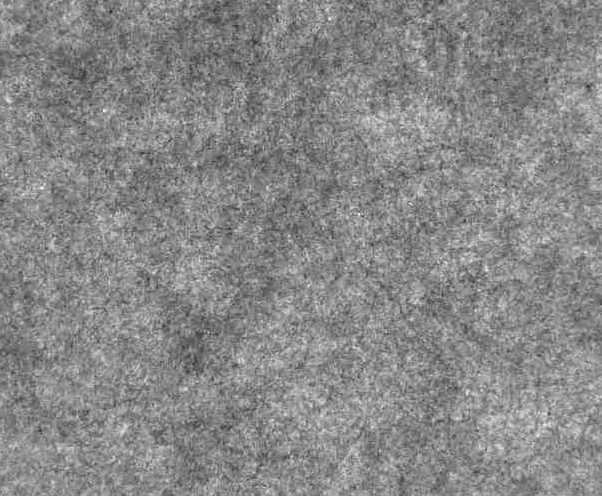}
	}
	
	\caption{\label{paperInLightSource} The surface and texture of the same area of a paper sheet as captured by a camera based on a) reflective and b) transmissive light.}
\end{figure}

Unlike previous works that measure the paper surface characteristics, we propose to fingerprint a paper sheet based on measuring the paper texture patterns. We capture the texture by putting a light source on one side of the paper and using a commodity camera to take a photograph on the other side. This is intuitively based on the common observation that putting a paper sheet under light will immediately reveal rich irregular textural patterns visible even to the naked eye. 
Figure~\ref{paperInLightSource} shows the difference between photos taken of the paper surface (based on reflective light) and of the paper texture (based on transmissive light). 



%

\textbf{Contributions.}
The main contributions of this paper are as follows:
\begin{itemize}
	\item We revisit paper fingerprinting and propose to use the textural patterns revealed by passing light through a paper sheet as a reliable source for extracting a fingerprint, as opposed to previous measures which are based on paper surface imperfections. 
	
	\item We design an efficient paper fingerprinting algorithm based on error correction and image processing techniques, and carry out experiments to show that our method can be used to efficiently extract a reliable and unique fingerprint using a photo taken by an off-the-shelf camera. Our proposed method is feasible and inexpensive to deploy in practice. 
	
	\item We conduct further experiments to demonstrate that our method is robust against: (a) non-ideal photo capturing settings such as when the paper is rotated and the light source is changed, and (b) non-ideal paper handling situations such as crumpling, soaking, heating and scribbling on the surface. 
	
\end{itemize}


\section{Paper Texture}
\label{paper}
When light falls on an object, it is partly absorbed, partly reflected, and partly transmitted, and paper is no exception. Absorption occurs based on the resonance principle: the energy of the light waves of a specific frequency is absorbed and transformed into kinetic energy by electrons of the same frequency. The part that is not absorbed, is either reflected or transmitted depending on how opaque (or conversely transparent) the paper is. 

Different types of paper behave differently in terms of how much light they absorb, reflect or transmit. This behaviour depends, among other factors, on pulp material, density, thickness and coating substances. Opacity, as defined by the ISO~2471 standard~\cite{iso}, can be seen as an indicator of how much light is impeded from transmitting through the paper, with the opacity of 100\% defined for fully opaque papers. Typical office printing paper, with grammage between 75 to 105~$g/m^2$, has opacity between 86\% to 94\%.  To put this in perspective, opacity for newsprint paper (typical grammage: 40--49~$g/m^2$) is in the range 90--94\% and for tracing paper (typical grammage: 60--110~$g/m^2$) is in the range 24--40\%~\cite{paperOpacity}. These values suggest that a considerable proportion of light transmits through such paper, which forms the basis of our proposal to fingerprint paper based on its textural patterns.

Intuitively, the textural patterns created and stabilized throughout the paper in the process of manufacturing can provide a promising source for paper fingerprinting. These patterns are naturally occurring and appear random. Moreover, they are embedded within the bonded structure of the paper and hence are relatively well-protected against manual handling of paper. They are generated as a result of the wooden particles randomly interleaved during the manufacturing process. Finally, once in the finished product, the randomly interleaved wooden particles can not be altered without damaging the paper, hence making any tampering act evident. 


To capture the embedded textural patterns of paper and subsequently extract a fingerprint, we limit ourselves to a single photo taken by a commodity camera. This makes our solution more practical and quicker than the previous proposal \cite{clarksonfingerprint} that has to take multiple scans (on paper surface) from four different angles in order to compute a fingerprint. We note that a single photo is feasible in our case because the paper texture contains richer features than the paper surface, such as the thickness of the overlaying wooden particles, randomly distributed impurities, and different embedded materials with varying opacities. In the rest of the paper, we conduct experiments to show that we can reliably extract a paper fingerprint from the textural patterns. 



{\bf Applications.} 
A vast number of official and legal documents, certificates, official receipts and invoices are printed on regular office paper (sometimes with watermarks, holograms or other security measures), thermal paper, or other types of paper. A property that the majority of these types of paper have in common is that they are not completely opaque. This means that a considerable amount of light passes through them. Furthermore, embedded irregular textural patterns as a natural result of the manufacturing process seem to be a universal property of all these different types of paper. Consequently, there is considerable potential for exploiting paper fingerprints extracted from embedded textural patterns in order to validate the authenticity of such official and legal documents.



\section{Texture Analysis}
\label{texture}
\begin{figure*}
	\centering
	\includegraphics[trim=100 500 0 0,clip,scale=0.6]{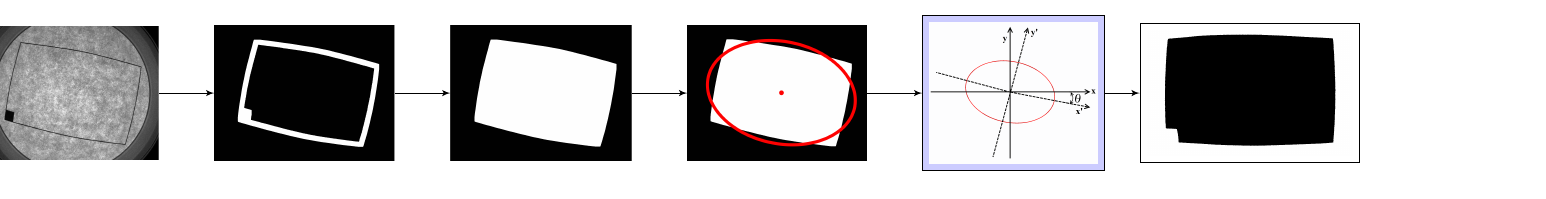}
	\caption{\label{rotation} Step-by-step rotation recognition process in the preparation phase. The last step produces a mask that distinguishes the pixels containing reliable information suitable for feature extraction (black region) from the pixels containing unreliable information (white region).}
\end{figure*}

In this section we discuss a high level description of our proposed method for capturing paper textural patterns and extracting a reliable and unique paper fingerprint from those patterns. 
To be able to capture paper textural patterns, we take a digital photograph of the paper sheet through which light is projected. 
Then, we need to perform a series of preparation operations such as aligning and resizing of the original image. 
Afterwards, in the texture analysis phase, we utilize 2-D Gabor filter~\cite{daugmanGabor} to extract textural information from the captured image. 
Subsequently, we propose a simple paper fingerprint extraction method that generates a binary string, the paper fingerprint. 
Once paper fingerprints are in the binary string format, they can be compared using well-known methods, such as computing the fractional Hamming distance between any two paper fingerprints. 
In the following, we give more details about the preparation phase, Gabor transform, the fingerprint generation method, and the fingerprint comparison method based on fractional Hamming distance. 
Further implementation details and settings of our experiments will be discussed in Section~\ref{evaluation}. 

\subsection{Preparation Phase}
The preparation phase consists of operations of identifying the designated area of the photo which is to be used for fingerprint extraction and aligning the image in terms of movement and rotation. 
To indicate the fingerprinting area, we print a small rectangular box on the paper sheet. In addition, we print a filled square on the bottom left of the box, to allow automatic alignment by our implementation. 

As shown in Figure~\ref{rotation}, aligning the rotation of the image involves several steps. First, we start with a photo of the fingerprinting area. The photo is converted into grey scale. The printed region (the rectangular box and the filled square) can be easily identified by applying a grey-scale threshold. This threshold is computed by the Otsu method~\cite{otsu}, which chooses the threshold in a such way to minimize the interclass variance of black and white pixels. We have applied the same approach for both reflection and transmission analyses. We observe that the borders in both reflection and transmission samples were recognized correctly using this technique. The result is a binary image: ``0'' for black and ``1'' for white. This simple thresholding may also produce some ``noise'' scattered around the image, but they can be easily removed based on area. To ensure the borders of the printed rectangle are connected, we draw a convex hull of the outer pixels to form a connected shape. This process also identifies artefacts, e.g., caused by pen scribbling as we will test in the robustness experiments. The pixel positions of identified artefacts are defined in a mask function, which we explain below.

Once the printed rectangle is identified, we fill up the region within the rectangular border with the binary value `1' (white). We identify the centre of mass of the rectangular object based on computing the first-order moment~\cite{imageRotation} and use that as the new origin of the Cartesian coordinate system. This corrects any misalignment due to paper movement. 




Then, we need to correct any misalignment caused by rotation. 
This is based on computing second-order moments~\cite{imageRotation} in the new Cartesian coordinate system. 
Let $B(x,y)$ denote the binary 2D object in Cartesian coordinates representing the recognized rectangular box area. 
There are three second-order moments as follows: 
\begin{align*}
u_{20} &= \iint x^2\,B(x,y)\,dx\,dy\ \\ 
u_{11} &= \iint x\,y\,B(x,y)\,dx\,dy\ \\ 
u_{02} &= \iint y^2\,B(x,y)\,dx\,dy\
\end{align*}
The rotation of the binary 2D object $B(x,y)$ can now be calculated as follows: 
\begin{equation}
\label{orientationEq}
\theta = \frac{1}{2} \, \tan^{-1}\left ( \frac{2 \, u_{11}}{(u_{20}-u_{02})+\sqrt{(u_{02}-u_{20})^2+4 \, u_{11}^2}} \right )
\end{equation}
The above formula -- based on a method originally proposed by Teague~\cite{imageRotation} -- calculates the angle between the $x$ axis and the major axis of an ellipse that has equal second moments to the recognized rectangular box. It gives us the counter-clockwise rotation of the object with respect to the horizon.  
After $\theta$ is calculated, the image can be rotated accordingly. 

It is worth noting that 
in the captured image, the borders of the rectangles are slightly curved rather than being straight due to lens artefact. This slight curvature does not affect our alignment algorithm.
We use the raw bitmap image acquired from the camera instead of the processed jpeg image. This raw image is stored separately in the camera in the ``.rw2'' format and contains the raw information captured by the camera sensor without any processing. 



After rotation is corrected, the image is delimited to the lowest and highest $x$ and $y$ values of the coordinates of the pixels inside the recognized rectangular box. This image is denoted by $I(x,y)$. Meanwhile, the mask for the image is calculated as $M(x,y)$. This mask is a binary vector with the same dimensions as $I(x,y)$ with the value `0' indicating the corresponding pixel in $I(x,y)$ to be masked out from the Hamming distance computation. In general, two categories of pixels are chosen to be masked out in our procedure. The first is the pixels with the intensity greater than the threshold computed by the Otsu method~\cite{otsu} and not considered as ``scattered noise'' in the border recognition phase. These include the printed rectangle, the filled square inside the box and any artefacts such as random pen scribbling. 
The second is the pixels outside the recognized box including all the edges in the picture. See the last diagram in Fig.~\ref{rotation} for an illustration. These pixels are considered to contain unreliable information. They are identified as `0' in a binary mask vector (similar to the identification of eyelids and eyelashes in iris recognition~\cite{daughmanIRIS2}) and will be excluded in the subsequent Hamming distance comparison process. 

%
%

\subsection{Gabor Filter}
Gabor filters are mainly used for edge detection in image processing. 
Besides, they have been found to perform efficiently in texture discrimination. 
Gabor filters are able to extract both coherent and incoherent characteristics of textural patterns~\cite{daughmanIRIS1}. Coherent properties are the patterns which remain unchanged between snapshots of the same sample while incoherent ones refer to the patterns which change between snapshots of different samples. 
The two dimensional Gabor wavelets are popular in biometric recognition problems such as iris recognition~\cite{daughmanIRIS2}, fingerprint recognition~\cite{gaborFingerprint} and face recognition~\cite{gaborFace}. 
A Gabor filter's impulse response is basically that of a Gaussian filter modulated by a sinusoidal wave. Consequently, Gabor filters capture features in both the frequency and spatial domains. Generally speaking, a Gabor filter would consider the frequency of a pattern (``what'') as well as the two-dimensional (2D) position of the pattern (``where'')~\cite{daughmanIRIS1}. 
Let $\exp$ be the natural exponential function. 
The 2D Gabor wavelet is calculated as follows using Cartesian coordinates: 

\begin{align}
\label{gaborEq}
& G(x,y) = \dfrac{f^2}{\pi \eta \gamma} 
\cdot \exp \left( \dfrac{ \eta^2 {x}'^2 + \gamma^2 {y}'^2 }{ 2 \sigma^2 } \right) 
\cdot \exp \left( 2 \pi i f {x}' \right) \\
& \text{for } \quad {x}'=x \, cos(\theta)+ y \, sin(\theta) \quad 
\text{and } \quad {y}'= -x \, sin(\theta) + y \, cos(\theta) \nonumber
\end{align}
where $f$ is the frequency of the sinusoidal wave, 
$\eta$ and $\gamma$ are constant factors that together determine the spatial ellipticity of the Gabor wavelet, 
$\theta$ represents the orientation of the ellipticity, and 
$\sigma$ is the standard deviation of the Gaussian envelope. 

Depending on the frequency of the sinusoidal wave and the orientation of their ellipticity, Gabor filters are capable of discriminating different textural characteristics. 
Usually, Gabor filters with a range of different frequencies, known as \emph{scales}, and a range of different orientations are applied to find out the best combination of scale and orientation for a specific texture analysis problem. 
For a fixed maximum frequency $f_{\max}$ and a maximum of $U$ scales, each scale index $u$ defines the frequency $f$ used in Equation~\ref{gaborEq} as follows: 
\begin{equation}
\label{scale}
\forall u \in \left \{ 1,2, \ldots ,U \right \} : f=\frac{f_{\max}}{\sqrt{2}^{u-1}}
\end{equation}
For a maximum of $V$ orientations, we consider $V$ angles equally distributed from 0 to $\pi$. 
Each orientation index $v$ defines the orientation $\theta$ used in Equation~\ref{gaborEq} as follows: 
\begin{equation}
\label{orientation}
\forall v \in \left \{ 1,2, \ldots ,V \right \}: \theta=\frac{v-1}{V} \, \pi
\end{equation}

%

We apply a Gabor filter to grey-scale images. 
Let $I(x,y)$ represent the grey-scale image using Cartesian coordinates. 
The result of the application of Gabor filter $G(x,y)$ is simply the 2D convolution of $I$ and $G$ as follows: 
\begin{align*}
\label{convolution}
C(x,y) &= I(x,y) \ast G(x,y) 
= \iint I(x,y) \ G(x-\eta ,y-\xi ) \  d\eta \ d\xi
\end{align*}
The result $C(x,y)$ is a complex number for each $x$ and $y$. 
$C(x,y)$ can be alternatively viewed as a matrix with the discrete values of $x$ and $y$ mapped to the columns and rows. 
Throughout the paper, we use functions defined over Cartesian coordinates and matrices interchangeably. 

\subsection{Fingerprint Generation}
\label{FPG}
\begin{figure}
	\centering
	\includegraphics[scale=1]{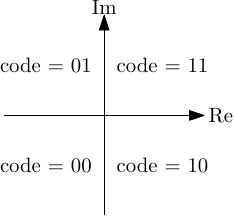}
	\caption{\label{unitCircle} Gray code for a complex value $m_{ij}=a+bi$ in the complex plain.}
\end{figure}

Our fingerprint generation method takes the output of a Gabor filter and produces a binary string. 
Let the element located in row $j$ and column $k$ of the matrix $C(x,y)$ be $m_{jk}=a+bi$. 
We define a 2-bit Gray code based on which quarter of the complex plane the element $m_{jk}=a+bi$ falls in (see Figure~\ref{unitCircle}). For example, 
when $a$ and $b$ are both positive, the encoded value will be $11$. Thus, every element in the matrix is replaced by two bits. 
The result is a binary string which we call the paper fingerprint.


\subsection{Fractional Hamming Distance}
After paper fingerprints are generated, fractional Hamming distance between any two fingerprints can be used to compare them. 
Hamming distance is simply the number of positions in which the bits disagree between two fingerprints. This is a classical bit error rate (BER) metric in communication.
Fractional Hamming distance is the normalized version, resulting a value between 0 and 1. 
Usually masking is used to discard the effect of irrelevant bits in a fingerprint. 
For each fingerprint, a mask is defined as a binary string of the same length in which bits corresponding irrelevant positions are set to 0 and bits corresponding effective positions are set to 1. 
The masks are calculated in the preparation phase as discussed above. 
Given two fingerprints $f_1$ and $f_2$, and their corresponding masks $m_1$ and $m_2$, the fractional Hamming distance is calculated as follows: 
\begin{equation}
\label{maskedHamming}
HD(f_1,f_2,m_1,m_2)=
\frac{\left \|(f_1 \oplus f_2)\cap m_1 \cap m_2 \right \|}{\left \| m_1 \cap m_2 \right \| }
\end{equation} 
where $\oplus$ denotes the bitwise exclusive-OR (XOR) operation and $\cap$ denotes the bitwise AND operation. 
A relatively small fractional Hamming distance indicates that the two fingerprints are likely to belong to the same paper sheet, while a relatively large fractional Hamming distance (around 0.5) indicates that the two fingerprints are likely to belong to different paper sheets. 
In the rest of the paper, we simply use Hamming distance (or HD for short) to refer to fractional Hamming distance.

\section{Evaluation}
\label{evaluation}
In order to evaluate our suggested method for paper fingerprinting, we collected several  datasets in different situations. 
In this section, we first explain the parameter settings and experiment configurations under which we carried out our evaluations. 
Then, we provide the details of the evaluation framework that we use to assess the results of our experiments. 
In particular, we consider metrics used for evaluating the effectiveness of biometric systems as well as those used for evaluating the effectiveness of physical unclonable functions (PUFs), since paper fingerprints can be seen as both. 
Subsequently, we give results that justify the choices we had to make in terms of how we collect our datasets and settings we use for Gabor filter. 
Finally, we give the details of our main dataset collection and provide the results of our experiments, including evaluation of the proposed method against biometric and PUF metrics.

\subsection{Parameter Settings \& Experiment Configurations}
\label{Conf}
In order to obtain consistent fingerprints, we require that a relatively small but fixed part of a sheet of paper is used as a source of fingerprint extraction. 
We chose to print a rectangular box (37mm$\times$57mm) on the sheet to indicate this area. 
In addition, we printed a small filled square (5mm$\times$5mm) at the bottom left of the box (see Figure~\ref{envpic}). 
Using this small square, in our preparation phase our method can check that the rotation has been carried out correctly (distinguishing cases when the paper is placed upside-down or flipped). 

The original photos in our experiments are all 3456$\times$4608 pixels. 
After the preparation phase, we get a corrected and delimited image of variable size, ranging between around 2300$\times$3300 pixels to 2350$\times$3350 pixels. 
This image is then resized to a 640$\times$640 pixel image $I$ which is then given as input to Gabor filter. The rectangular size conversion is for the convenience of applying Gabor wavelets in the next stage to produce 2048 bits in the output (the same size as an iris code). We use a Gabor impulse response of size 100$\times$100 and the output of Gabor filter in our experiments, $C$, is a complex matrix of size 640$\times$640. 
This matrix is downsampled to one of size 32$\times$32, before being given as input to the fingerprint generation algorithm. This downsampling process is done by simply picking the elements in every 20th row and 20th column. 
Fingerprint generation replaces each complex value with two bits. 
Hence, the final paper fingerprint is a string of the size 2$\times$32$\times$32=2048 bits. 


We chose to downsample the output of the Gabor filter for two reasons. First, it makes the data storage more compact. With 2048 bits (256 bytes), we are able to store the fingerprint in a QR code as part of an authentication protocol (we will explain the protocol in more detail in Section \ref{security}). Second, adjacent pixels in the image are usually highly correlated. Hence, downsampling serves to break the correlation between bits. Through experiments, we found this simple downsampling technique was effective to produce reliable and unique fingerprints.

\begin{figure}
	\centering
	\subfigure[\label{camera} Camera and macro flash ring.]
	{
		\includegraphics[scale=0.047]{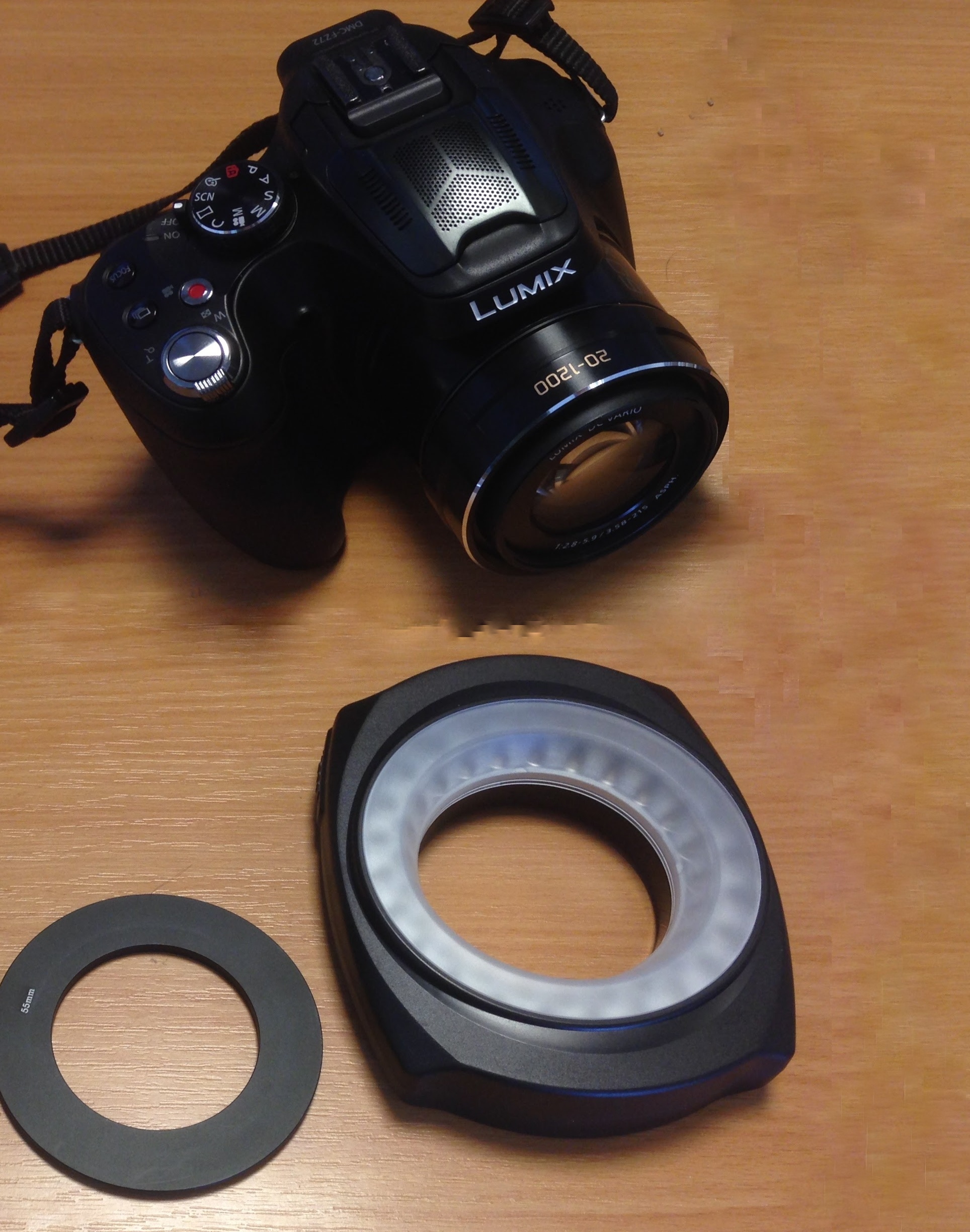}
	}
	\subfigure[\label{overhead} Overhead projector as light source.]
	{
		\includegraphics[scale=0.077]{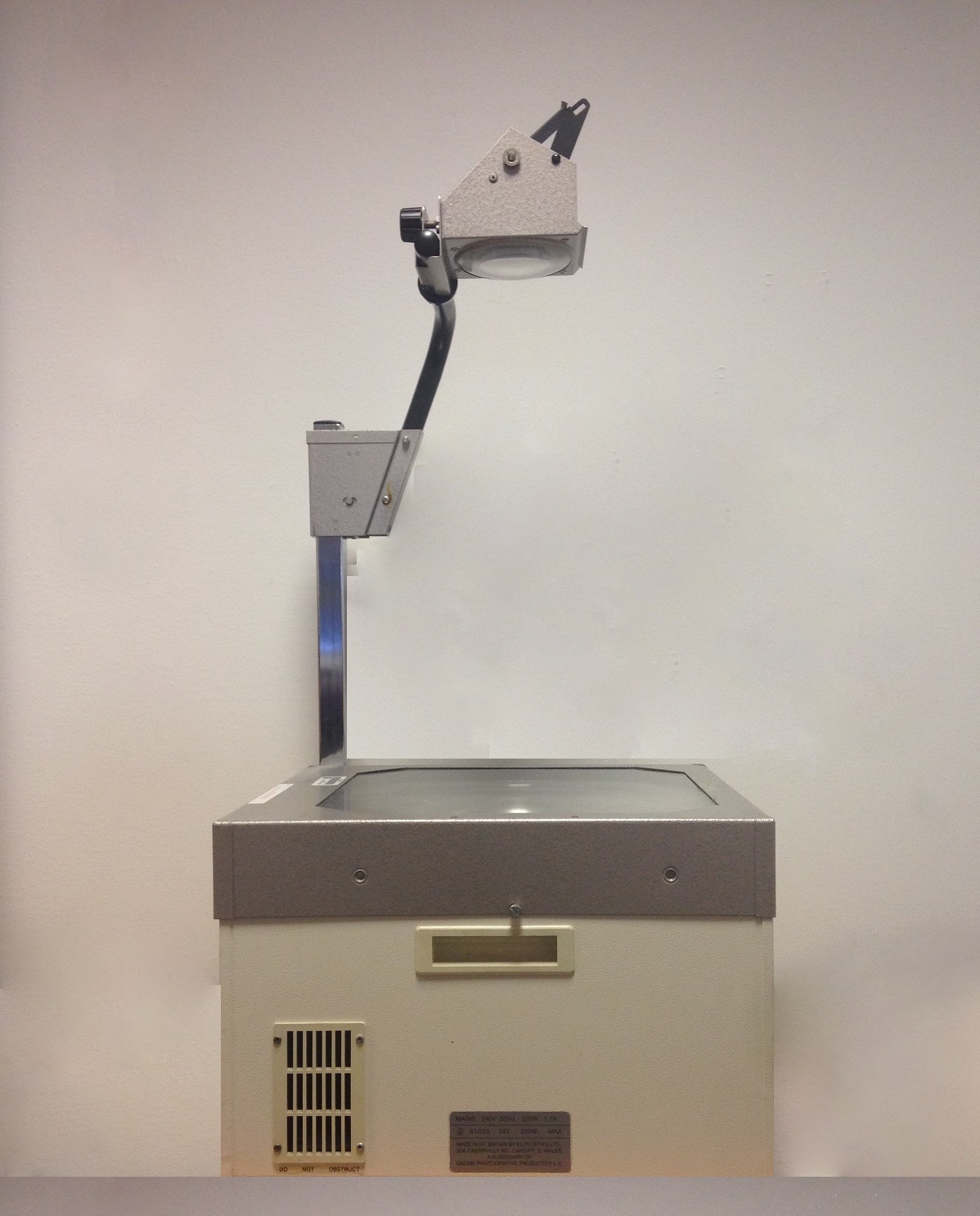}
	}
	\subfigure[\label{lightbox} Light box (tracing pad) as light source.]
	{
		\includegraphics[scale=0.138]{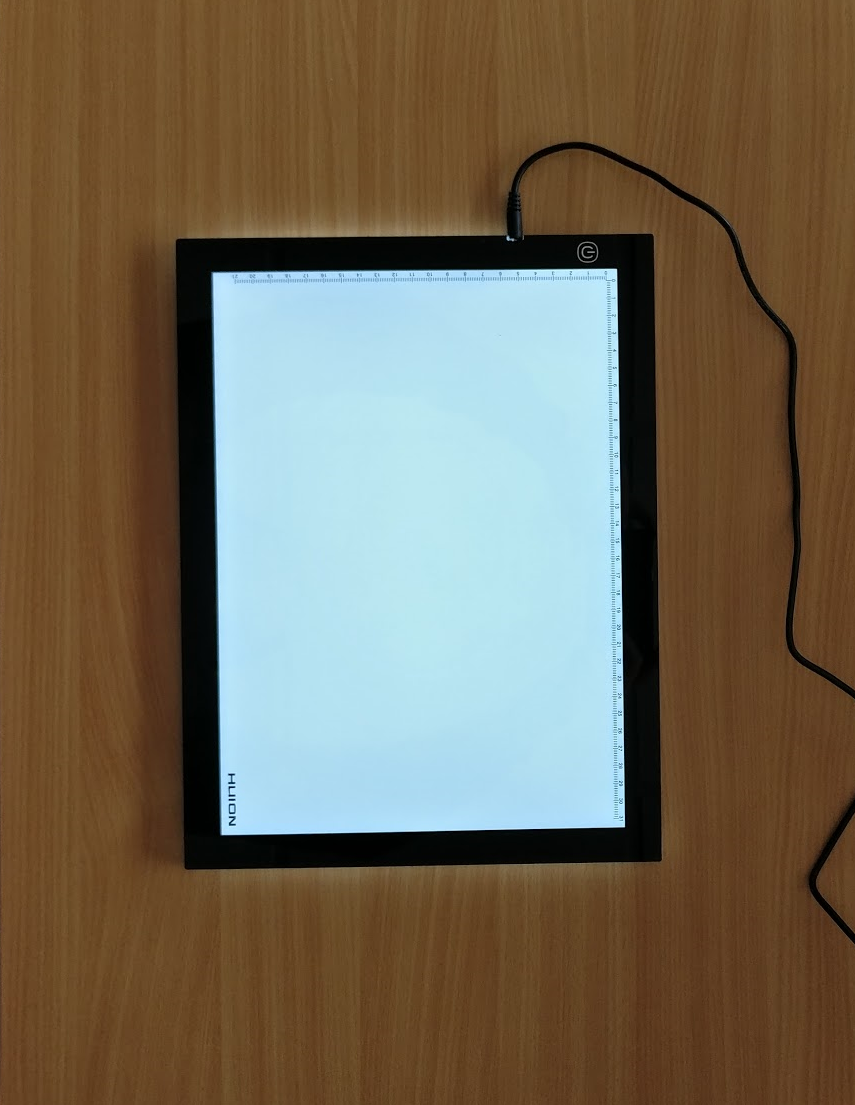}
	}
	\caption{\label{implementation} The equipment used in our experiments.}
\end{figure}

All images have been captured by a Panasonic DMC-FZ72 camera with a resolution of 16.1 Mega-pixels. 
We chose this camera for two main reasons: the ability to capture a photo in macro mode from a short distance (minimum 1 cm focus) and the ability to mount a macro flash ring over the lens. 
However, these characteristics are not unique to this specific camera and many other cameras available in the market provide the same characteristics. 
We mounted an off-the-shelf common macro flash ring on the camera lens, to maintain a constant distance between the lens and the paper surface where the texture is photographed. 
The camera and its accessories are shown in Figure \ref{camera}.
In our experiments, we do not use the flash of the macro flash ring; the light source is an ordinary office overhead projector as shown in Figure \ref{overhead}. 
The light that the overhead projector provides is intense and adjustable. Furthermore, it has a flat surface with constant distance from the light source. This allows us to put the paper on the surface and then the macro ring resting on top of it before the camera takes a photo of the paper texture. The use of the macro ring also serves to shield the effects of other ambient light sources (e.g., daylight, office lighting). In Section~\ref{lightBoxSource}, we will explain the effect of the light source by using an alternative source: a commodity light box (tracing pad) as shown in Figure~\ref{lightbox}. 

Our evaluations were performed on a PC with an Intel Core~i7-2600S CPU @ 2.80~GHz with 8~GB of memory. The operating system was 64-bit Windows~7 Enterprise and we used Matlab R2015a (64-bit) to develop our algorithms. 

\subsection{Evaluation Framework}
\label{framework}

Our work is closely related to the fields of biometrics and Physical Unclonable Functions (PUFs). Biometrics is the science of authenticating humans by measuring their unique characteristics and have a long history of research. A paper fingerprint works similar to biometrics, except that it measures unique characteristics of a physical object instead of a human being. Hence, common metrics that measure the error rate performance of a biometric system apply to our work too. On the other hand, paper fingerprints are related to Physical Unclonable Functions, which is a relatively new field starting from Pappu et~al.'s seminal paper published in \emph{Science} in 2002~\cite{pufMain}. Typically PUFs require a challenge and response dynamic, but according to the definition by Maes in~\cite{opticalPUF1}, paper can be regarded as a ``non-intrinsic'' PUF, i.e., a PUF that does not contain the circuitry to produce the response on its own. Hence, the same evaluation methods in PUF are also applicable to paper fingerprints. 

Because of the close relation to these two fields and their respective evaluation frameworks, we evaluate our method based on metrics used in both fields for a comprehensive analysis. We note that previous works either use biometric  or PUF metrics, however using both allows us to perform a meaningful comparison with related biometrics and PUF systems. It would be interesting to study the relationships between these metrics and define a unified framework that can be applied to evaluate both biometric and PUF systems. We leave this to future work.


In the following we give a brief description of these metrics. 
We discuss Hamming distance distributions, decidability, and recognition rates including false rejection and false acceptance rates in the former category of metrics. 
In the latter category, we consider uniformity and randomness in the space dimension, reliability and steadiness in the time dimension, and uniqueness and bit aliasing in the device dimension. 

{\bf 1) Biometric Metrics:} 
A biometric authentication problem is a specific case of a statistical decision problem in which one decides if two given biometric measurements belong to the same source or not. 
In order to provide necessary information about the effectiveness of such a biometric, the parameters of the so-called biometric decision landscape need to be specified~\cite{DaugmanLandscape}. 
If Hamming distance is used for comparison, as it is in our case, the distributions of Hamming distance for two groups of comparisons need to be determined: for comparisons between paper fingerprints originating from the \emph{same} paper sheet, and for comparisons between paper fingerprints originating from \emph{different} paper sheets. 
These are called \emph{same-group} and \emph{different-group} distributions, respectively. 

For an effective biometric, the same-group and different-group distributions should be well-separated. 
This makes the decision problem solvable. 
Let $\mu_1$ and $\mu_2$ denote the means, and $\sigma_1$ and $\sigma_2$ the standard deviations of the two distributions. 
Daugman defines the \emph{decidability} metric $d'$ as follows~\cite{irismain}: 
\begin{equation}
\label{decidability}
d' = \frac{ \left | \mu_1 - \mu_2 \right | }{ \sqrt{\frac{{\sigma_1}^2 + {\sigma_2}^2}{2}} }
\end{equation}
where $|\cdot|$ denotes absolute value. 
Decidability as defined above is indicative of how well-separated the two distributions are: the further and the more concentrated the distributions are, the higher will the decidability be. 
To give an idea about typical values, the decidability of iris recognition, a well-established and effective biometric method, is $d' \approx 14$ in an ideal measurement environment and $d' \approx 7$ in a non-ideal environment~\cite{irismain}. 


After determining the same-group and different-group distributions, one decides a threshold value situated between the two distributions. 
Subsequently, the decision on whether two reported biometrics belong to the same origin or not is then made by computing the Hamming distance between the two biometric samples and comparing it to the threshold. 
For an effective biometric, measurements from the same origin have relatively low Hamming distance and hence fall below the threshold, whereas measurements from different origins have relatively high Hamming distance and fall above the threshold. 
If the distributions are completely separated, the decision is correct all the time. 
However in practice usually there is some overlap between the two distributions. 
The proportion of biometrics from different origins falsely accepted as being from the same origin is known as the \emph{false acceptance rate (FAR)}. 
The proportion of biometrics from the same origin falsely rejected as being from different origins is known as the \emph{false rejection rate (FRR)}. 
For an effective biometric FAR and FRR should be low -- ideally zero. 

A widely used measure of effectiveness of a biometric is \emph{degrees of freedom (DoF)}. 
DoF is a measure of the combinatorial complexity of the biometric test, or in other words the number of bits in a biometric measurement that are independent~\cite{irismain}. 
Consider a biometric that provides degrees of freedom $N$, that is, $N$ independent and unpredictable bits. 
A comparison between two such biometrics from different origins can be modelled as the probability that a threshold number of $N$ independently chosen bits agree. 
Hence, the different-group distribution for such a biometric would follow the binomial distribution with mean $\mu=p$ and variance $\sigma^2=Np(1-p)$, where $p$ is the probability of single bit agreement. 
Hence, the degrees of freedom for a biometric with a different-group distribution that follows a binomial distribution with mean $\mu$ and variance $\sigma^2$ can be calculated as follows: 
\begin{equation}
\label{eq:dof}
N = \dfrac{\mu (1-\mu)}{\sigma^2}
\end{equation}

{\bf 2) PUF Metrics.} 
Paper fingerprinting can be seen as an optical physical unclonable function (PUF)~\cite{pufMain}, as pointed out in the literature~\cite{opticalPUF1,opticalPUF2}. However, previous works on paper fingerprinting did not evaluate their results in this context. 
We believe that evaluating our results against established PUF metrics provides further information about the effectiveness of our method and helps put our results in perspective within PUF literature. 



We follow the unified framework put forward by Maiti et al.~\cite{statPUF}. 
This framework provides metrics to evaluate a PUF in three dimensions: space, time, and device. 
In our case, PUFs are the paper fingerprints, and devices are the different paper sheets. 
Each of these dimensions quantifies a specific quality of a fingerprint: 
the space dimension analyses the overall variations of fingerprints, 
the time dimension indicates same-group consistency, and 
the device dimension discusses the different-group diversity of fingerprints. 

Before describing these dimensions, let us define the symbols we use in this framework. 
Here we consider \emph{effective fingerprints}, denoted by $r$. 
The effective fingerprint is the result of applying the appropriate mask over the original fingerprint $f$. 
We use the following parameters: 
$L$ is the number of bits in each fingerprint (2048 in our setting). 
$T$ refers to the number of samples taken from each paper sheet in a dataset (e.g., in the our benchmark dataset $T=10$). 
$N$ is the total number of paper sheets involved in a dataset (e.g., in the our benchmark dataset $N=100$). 
We use the following indices accordingly: 
$n$ denotes the paper sheet number within different sheets, 
$t$ represents the sample number within the samples from the same paper sheet, and 
$l$ shows $l$-th bit in the effective fingerprint. 

\subsubsection{Space Dimension} 
This dimension is concerned with bit variations with respect to the locations of the bits in fingerprints. 
Metrics in this dimension evaluate the overall inter-sheet behaviour of fingerprints. 

\begin{itemize}
	\item {\it Uniformity:} 
	This metric shows how uniform 0s and 1s are in a fingerprint. 
	The ideal value for this metric is 0.5. 
	Uniformity of the fingerprint from the $t$-th sample and $n$-th sheet is calculated as follows: 
	
	\begin{equation}
	\label{uniformity}
	\mathrm{Uniformity}(n,t) = \frac{1}{L}\sum_{l=1}^{L}r_{n,t,l}
	\end{equation}

	\item {\it Randomness:} 
	This metric indicates the average randomness of the bits in the fingerprints generated from several acquisitions from a sheet. 
	The ideal value for this metric is 1.
	Randomness of the fingerprint bits generated from the $n$-th sheet is calculated as follows: 
	
	\begin{align}
	\label{randomness}
	\mathrm{Randomness}(n) &= -\mathrm{log}_2 \, \mathrm{max}(p_n,1-p_n),
	\\ 
	\text{where}\quad p_n &= \frac{1}{TL}\sum_{t=1}^{T}\sum_{l=1}^{L}r_{n,t,l} \nonumber 
	\end{align}
\end{itemize}

\subsubsection{Time Dimension} 
This dimension is concerned with fingerprint variations within multiple samples. 
Metrics in this dimension evaluate the overall intra-sheet persistence of fingerprints within multiple samples. 

\begin{itemize}
	\item {\it Reliability:} 
	This metric shows how consistently fingerprints are reproduced by the same sheet. 
	The ideal value for this metric is 1.
	Reliability of the fingerprints generated from the $n$-th sheet is calculated as follows:
	\begin{equation}
	\label{reliability}
	\mathrm{Reliability}(n) =  
	1 - \frac{2}{T (T-1) L} \sum_{t_1=1}^{T-1} \sum_{t_2=t_1+1}^{T} \sum_{l=1}^{L}
	(r_{n,t_1,l} \oplus r_{n,t_2,l})
	\end{equation}
	
	\item {\it Steadiness:} 
	This metric indicates the bias of individual fingerprint bits on average for a sheet. 
	The ideal value for this metric is 1.
	Steadiness of the fingerprints generated from the $n$-th sheet is calculated as follows:
	
	\begin{align}
	\label{steadiness}
	\mathrm{Steadiness}(n) &= 1+ \frac{1}{L} 
	\sum_{l=1}^{L}\mathrm{log}_2 \, \mathrm{max}(p_{n,l},1-p_{n,l})
	\\ 
	\text{where}\quad p_{n,l} &= \frac{1}{T}\sum_{t=1}^{T}r_{n,t,l} \nonumber 
	\end{align}
	
\end{itemize}

\subsubsection{Device Dimension} 
This dimension is concerned with fingerprint variations between multiple sheets. 
Metrics in this dimension evaluate the overall inter-sheet distinguishability of fingerprints.  

\begin{itemize}
	\item {\it Uniqueness:} 
	This metric represents how distinguishable a sheet is within a group of sheets. 
	The ideal value for this metric is 0.5.	
	Uniqueness of the fingerprints generated from the $n$-th sheet is calculated as follows:
	
	
	
	\begin{equation}
	\label{uniqueness}
	\mathrm{Uniqueness}(n) =   
	\frac{1}{T^2 L (N-1)} \cdot 
	\sum_{t=1}^{T} \sum_{\substack{n'=1\\ n'\neq n}}^{N} 
	\sum_{t'=1}^{T} \sum_{l=1}^{L} (r_{n,t,l} \oplus r_{n',t',l})
	\end{equation}
	
	\item {\it Bit-Aliasing:} 
	This metric indicates how likely different sheets are to produce identical fingerprint bits. 
	The ideal value for this metric is 0.5.
	Bit-aliasing of the $l$-th bit of the fingerprints generated from a dataset is calculated as follows:
	
	\begin{equation}
	\label{bit_aliasing}
	\mathrm{Bit}\text{-}\mathrm{Aliasing}(l)=\frac{1}{N T}\sum_{n=1}^{N} \sum_{t=1}^{T} r_{n,t,l}
	\end{equation}
	
\end{itemize}

%

\begin{figure}
	\centering
	\subfigure[Transmission]
	{
		\includegraphics[scale=0.25]{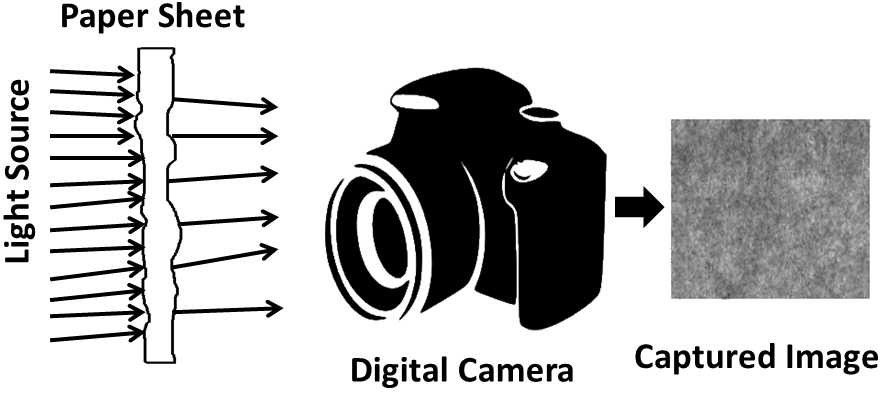}
	}
	\subfigure[Reflection]
	{
		\includegraphics[scale=0.25]{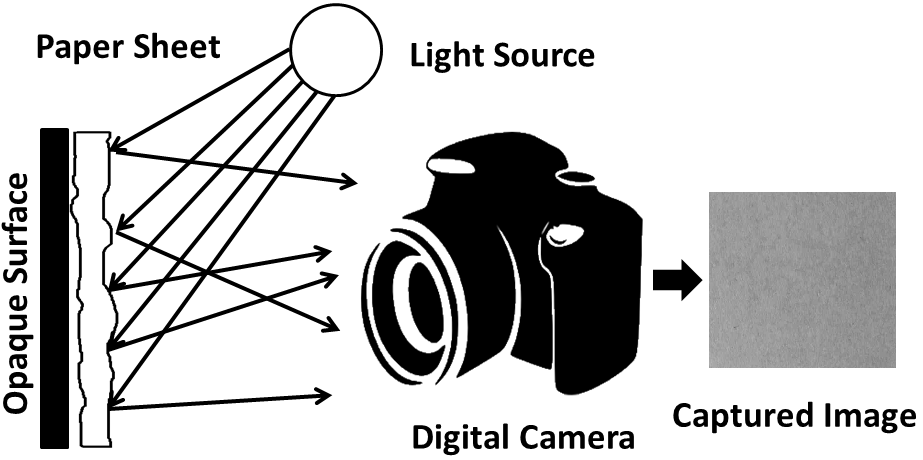}
	}
	
	\caption{\label{lightCamera} Capturing a photo, in case of (a) transmission, and (b) reflection, using the same digital camera and light source.}
\end{figure}

\subsection{Reflection vs.\ Transmission}
As discussed before, the main motivation of our work is to capture paper textural patterns and efficiently extract unique paper fingerprints from such patterns using an off-the-shelf camera. 
By contract, previous works~\cite{paperPUF,clarksonfingerprint,paperspeckle} extract paper fingerprints from the paper surface. Our hypothesis is that textural patterns revealed by the transmissive light contain richer features than the paper surface shown by the reflective light. To verify this hypothesis, we set up an experiment to investigate the difference between the two patterns. 


%

We set up the paper photographing in two settings: one with the light source on the same side of the paper and the other with the light source on the opposite side of the paper (see Figure~\ref{lightCamera}). In the former, we put an opaque object behind the paper, so only the paper surface is photographed based on the reflective light. We selected 10 common A4 (210$\times$297~mm) paper sheets with grammage 80~~$g/m^2$. We took 10 photos of each sheet in each of the two settings. We used a common overhead projector as our light source. 
We tried to reduce the effect of any ambient light by setting our data collection environment in a dark room. This data collection resulted in two datasets: a 100-sample dataset (10 sheets with 10 samples for each sheet) for surface measurements, and a 100-sample dataset (10 sheets with 10 sample for each sheet) for textural measurements. 

\begin{figure}
	\centering
		\includegraphics[scale=0.25]{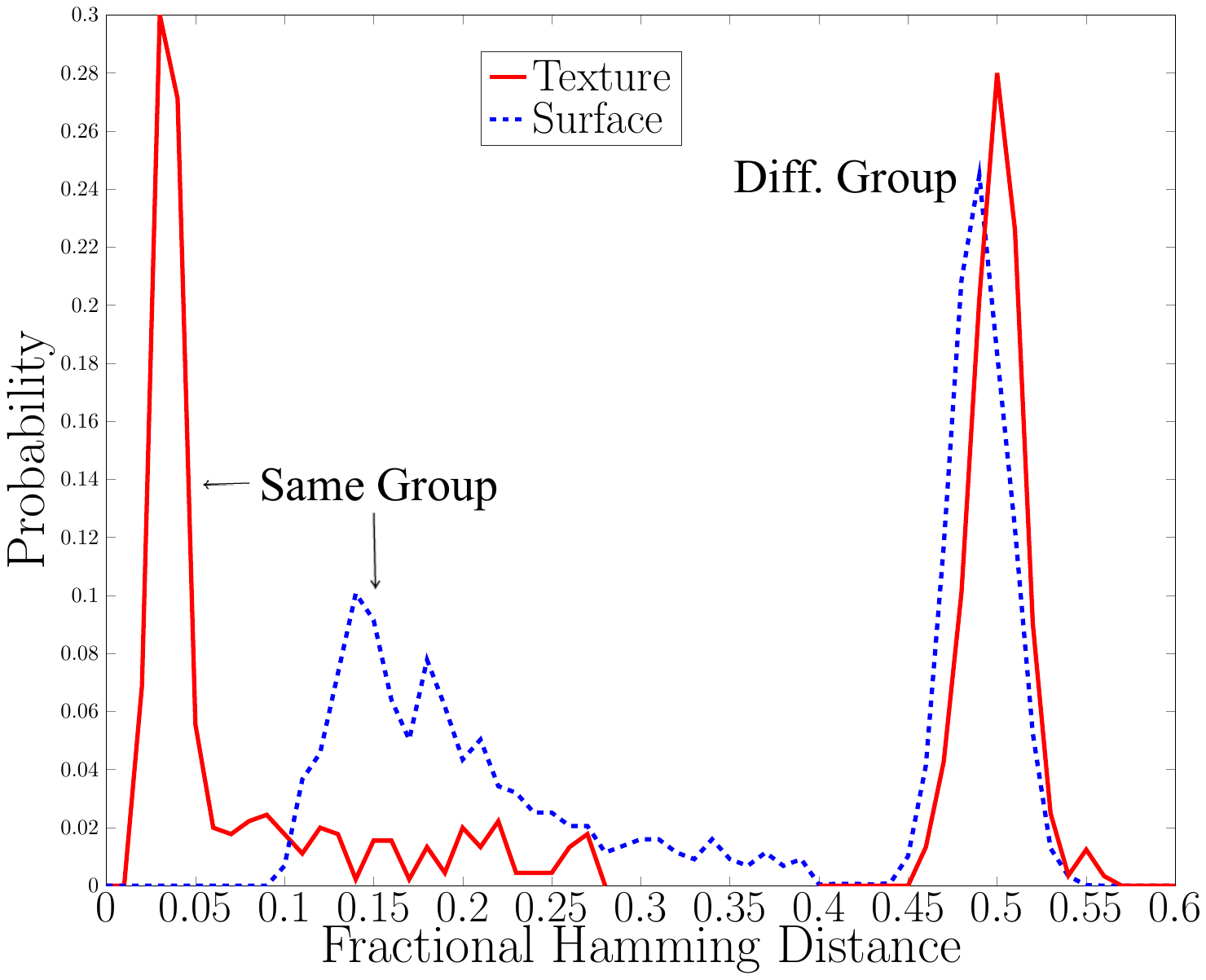}
	\caption{\label{reflectionvstransmission} Hamming distance distributions for surface and texture.}
\end{figure}

After the data collection, we performed our fingerprint extraction algorithm (as discussed in Section~\ref{texture}) for both datasets. Figure~\ref{reflectionvstransmission} shows the Hamming distance distributions for the two cases. Each diagram depicts four distributions: for each case i.e., surface and texture, there is one curve, concentrated around lower values of Hamming distance, showing the distribution of Hamming distance between pairs of fingerprints of the \emph{same} paper sheet, and a second curve, concentrated around a Hamming distance value of about 0.5, showing the distribution of Hamming distance between pairs of fingerprints of \emph{different} paper sheets. 

Ideally, for effective fingerprint recognition, we want the ``same-group'' and ``different-group'' distributions to be as separate as possible, since then we can easily decide on a threshold and consider any two fingerprints with a Hamming distance below that threshold to belong to the same paper sheet, and consider any two fingerprints with a Hamming distance above that threshold to belong to different paper sheets. 

As can be seen in Figure~\ref{reflectionvstransmission}, the two distributions, i.e., ``same-group'' and ``different-group'', are well-separated in the case of texture, but less so in the case of surface. In fact, in the case of texture, the minimum Hamming distance for different comparisons is 0.46 and the maximum Hamming distance for similar comparisons is 0.27, which shows that there is no overlap between the two distributions. However, in the case of surface, the minimum Hamming distance for different comparisons is 0.44 and the maximum Hamming distance for similar comparisons is 0.48, which shows that there is some overlap between the two distributions, and hence false negative or false positive decisions are inevitable in this case. Indeed, decidability for the case of texture is around 20, but for the case of surface it is around 6. 
Furthermore, the number of degrees of freedom provided by the texture is slightly higher than that provided by the surface. 
These results support our hypothesis that the textural measurements through transmissive light contain more distinctive features than surface measurements based on reflective light, and hence can be used as a more reliable source for paper fingerprinting. 

We should stress that the hypothesis is tested using a specific image capturing condition in which only one snapshot is taken. One should not directly compare the results with Clarkson et al.'s 3D method~\cite{clarksonfingerprint}, which is carried out in a different test condition and involves taking four scans from four different angles on the paper surface. However, we believe a method that is based on taking a single snapshot is easier and quicker than those that require multiple measurements. 

\subsection{Determining Gabor Scale \& Orientation}
As discussed, Gabor filter can be configured with different scales and orientations. 
To find out the appropriate combination of scales and orientation for our method, we set up an initial experiment. 
We collected a dataset including two sub-datasets: 
the first one includes 20 samples from one paper sheet; 
the second one includes one sample from each of 20 paper sheets. 
These two sub-datasets constitute our same-group and different-group data, respectively. 
We applied Gabor filter for 8 orientations, indexed from 1 to 8, representing angles $0$, $\frac{\pi}{8}$, $\frac{\pi}{4}$, $\frac{3\pi}{8}$, $\frac{\pi}{2}$, $\frac{5\pi}{8}$, $\frac{3\pi}{4}$, and $\frac{7\pi}{8}$. 
Considering $f_{\max}=0.25$, we also considered multiple scales, indexed by integer values starting from scale 1. 
We used fixed values of $\eta=\gamma=\sqrt{2}$ and $\sigma=1$. 

Ideally, we would want the different-group distribution to be centred around 0.5 or a mean very close to 0.5. 
Our experiments show that for scales greater than 7, the mean of the different-group distribution falls below 0.45, which indicates undesirable bias on the binomial distributions (i.e., tossing a coin is no longer random in the a Bernoulli trial~\cite{daughmanIRIS2}). 
Therefore, in the following we limit the scope of our investigation to  scales from 1 to 7. 

Our calculations show that as the scale increases, the decidability of the distributions increases, but at the same time the number of the degrees of freedom the different-group distribution provides decreases. This is because the scale relates to the spatial frequency components of the Gabor filer -- the smaller the scale is, the more detailed the feature extraction is. When the scale is one, the finest detail of the paper texture is extracted, which leads to high degrees of freedom in the generated fingerprint. However, at this scale, the image processing is extremely sensitive to noise, which reduces the separation between the same-group and different-group histograms of Hamming distances. Increasing the scale results in a zooming-out effect. More correlations between bits are introduced, which reduces the degrees of freedom. But on the other hand, the feature extraction is more tolerant of noise. As a result, the same-paper and different-paper characteristics become more distinctive, which leads to a higher decidability.

\begin{figure*}
	\centering
	\subfigure[\label{scaleDecidability} Decidability for all orientations and scales.]
	{
		\includegraphics[scale=0.226]{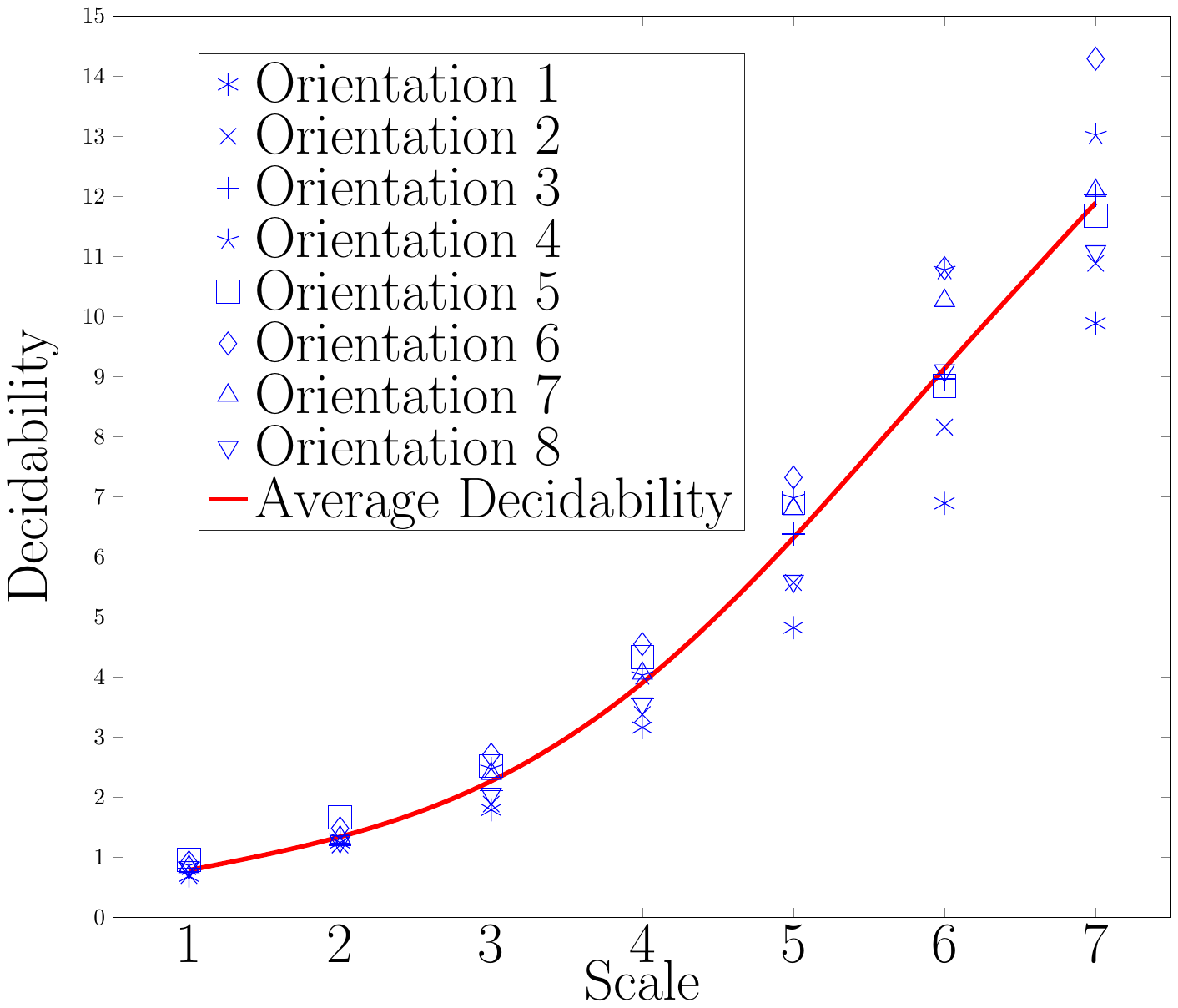}
	}
	\subfigure[\label{dofComparison} Degrees of freedom for all orientations and scales.]
	{
		\includegraphics[scale=0.226]{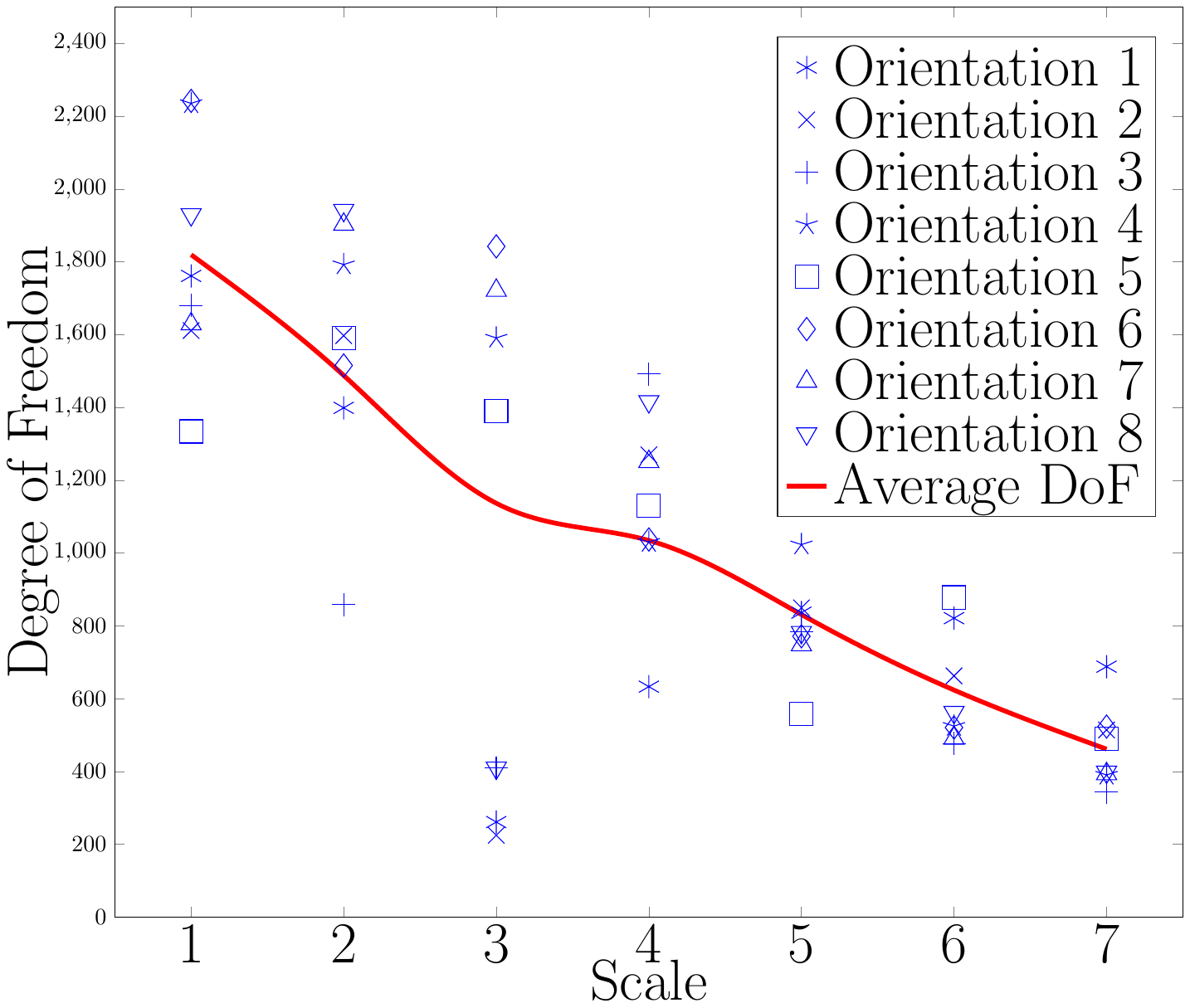}
	}
	\caption{Results of (a) decidability and (b) degrees of freedom in scales 1 to 7 and orientations 1 to 8.}
\end{figure*}

The results for decidability and degrees of freedom for orientations 1 to 8 and scales 1 to 7 are shown in Figures~\ref{scaleDecidability} and \ref{dofComparison}, respectively. 
Both figures also include a spline interpolation of average values of different orientation results within each scale to highlight the dominant trends. 
Therefore, there is an evident trade-off in choosing the scale and orientation. 
Too low a scale would not provide an acceptable decidability, while too high a scale would not provide a reasonable degree of freedom.
Through experiments, we find that the combination of scale 5 and orientation 7 provides a good trade-off between decidability and degrees of freedom. As we will explain later, this combination provides nearly perfect recognition rates. 
In the rest of the paper, we report all our findings based on this specific configuration of Gabor filter.



\subsection{The Benchmark Dataset}
\label{sets}
Our main dataset on which we report our evaluations is a set of 1000 samples collected by taking 10 photos of each of 100 different paper sheets to provide a good diversity. 
We use typical office paper sheets of size A4 (210mm $\times$ 297mm) with grammage of 80~$g/m^2$. 
All the sheets were from the same pack with the same brand. 
In all of the photos, camera settings including aperture and exposure time were kept constant. 
We tried to keep the paper sheets visually aligned for the different samples, and conducted separate experiments to evaluate the robustness of our algorithm against rotations (which we discuss in Section \ref{robustness}). 
We refer to the main dataset collected here under relatively stable conditions as the benchmark dataset. 


\subsection{Experiment Results}
\label{res}
In the following, we present the results of our experiments reporting the metrics introduced in Section~\ref{framework}. 
We also present the timing measurements for our method and provide a short discussion on its practicality. 
We provide comparison with existing work whenever the relevant metrics are reported in the literature. 

\begin{figure}
	\centering
	
	\includegraphics[scale=0.225]{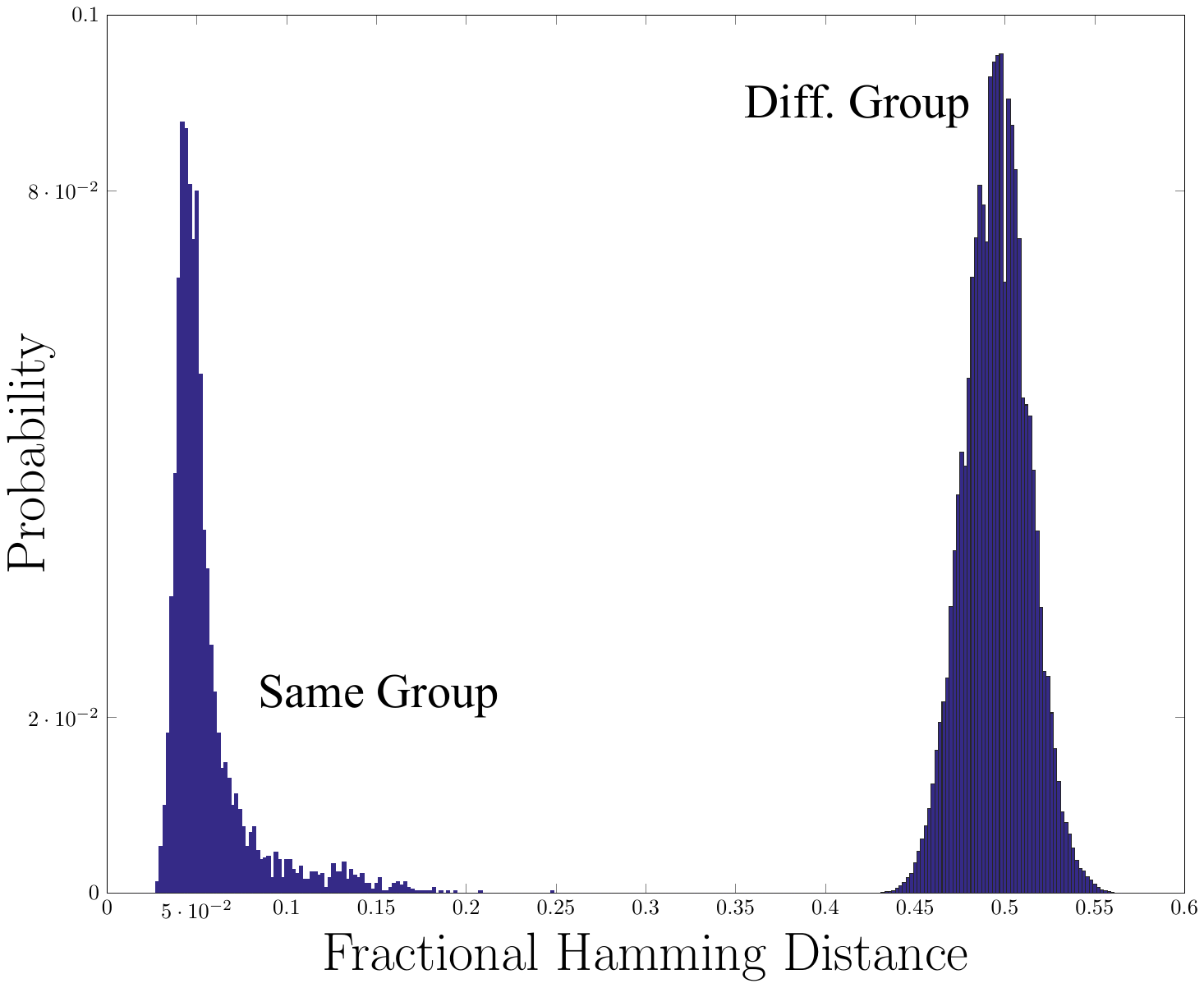}
	\caption{\label{hamming_distances} Hamming distance distributions in the benchmark dataset.}
\end{figure}

{\bf Biometric metrics. } 
We calculated the Hamming distance for all comparisons, consisting of same-group comparisons and different-group comparisons. 
There are a total of $\bigl(\begin{smallmatrix}1000\\2\end{smallmatrix}\bigr)= 499,500$ comparisons, of which $100 \cdot \bigl(\begin{smallmatrix}10\\2\end{smallmatrix}\bigr) = 4,500$ are same-group comparisons and $\frac{1000 \times 990}{2}=495,000$ are different-group comparisons. 
In Figure~\ref{hamming_distances} we show the distributions for the same-group and different-group Hamming distance values. 
Clearly, the two distributions are well-separated, which shows the effectiveness of our paper fingerprinting method. 
Indeed, the maximum same-group Hamming distance is 0.24, whereas the minimum different-group Hamming distance is 0.42, which shows that there is no overlap between the two distributions. 
Hence, any threshold between the above values would give us FAR and FRR of zero. 
As an example, we can choose the threshold to be 0.4, but this is adjustable. Detailed error rate performance will be reported in Section \ref{comparisonHamming}.


\begin{figure*}
	\centering
	
	\subfigure[Histogram of same-group HDs with  
	$\mu = 0.056$, $\sigma = 0.024$ ]
	{
		\includegraphics[scale=0.22]{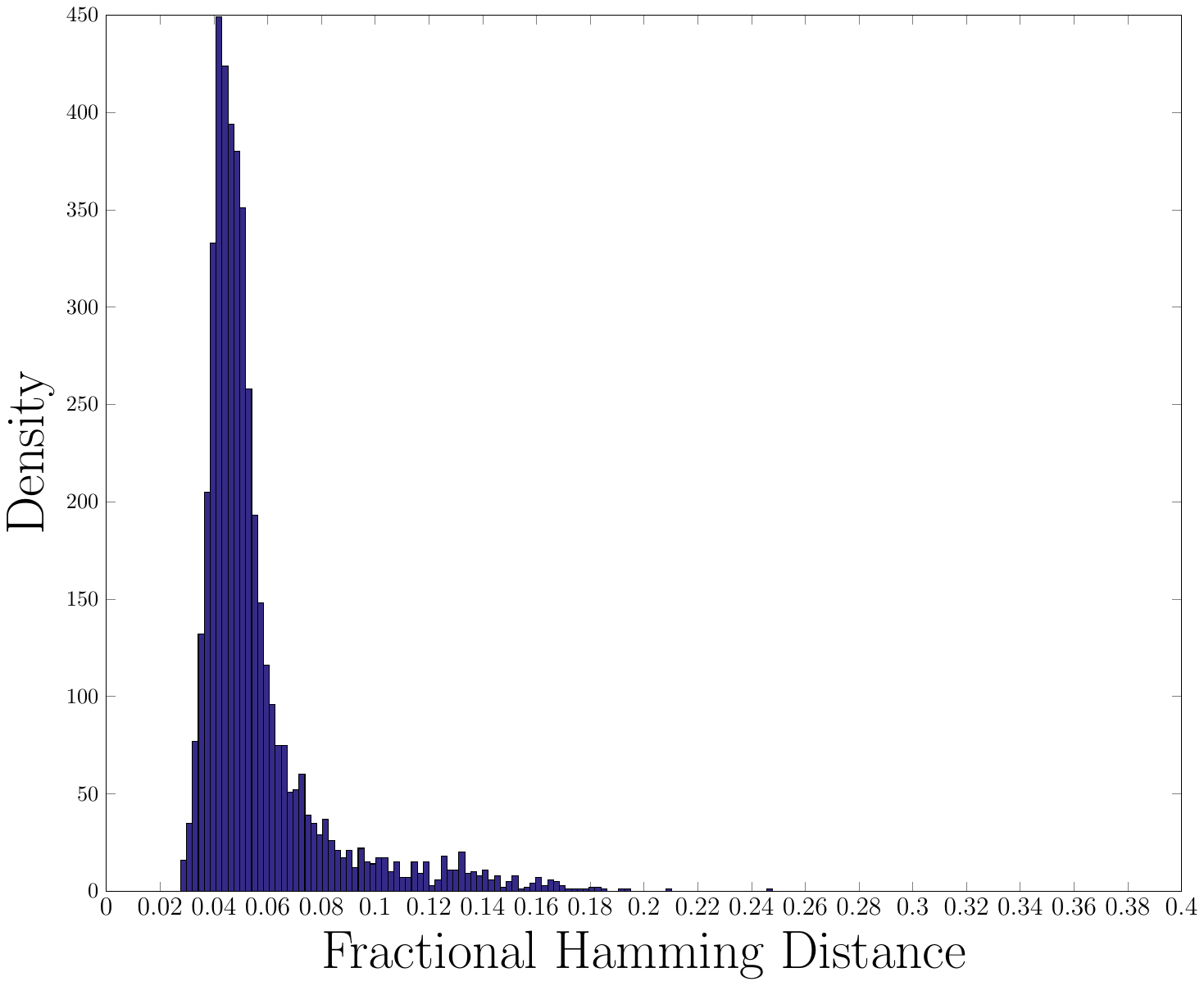}
	}
	\quad
	\subfigure[Histogram of different-group HDs with a 
	binomial curve with $N=807$, $\mu = 0.495$, $\sigma = 0.018$]
	{
		\includegraphics[scale=0.22]{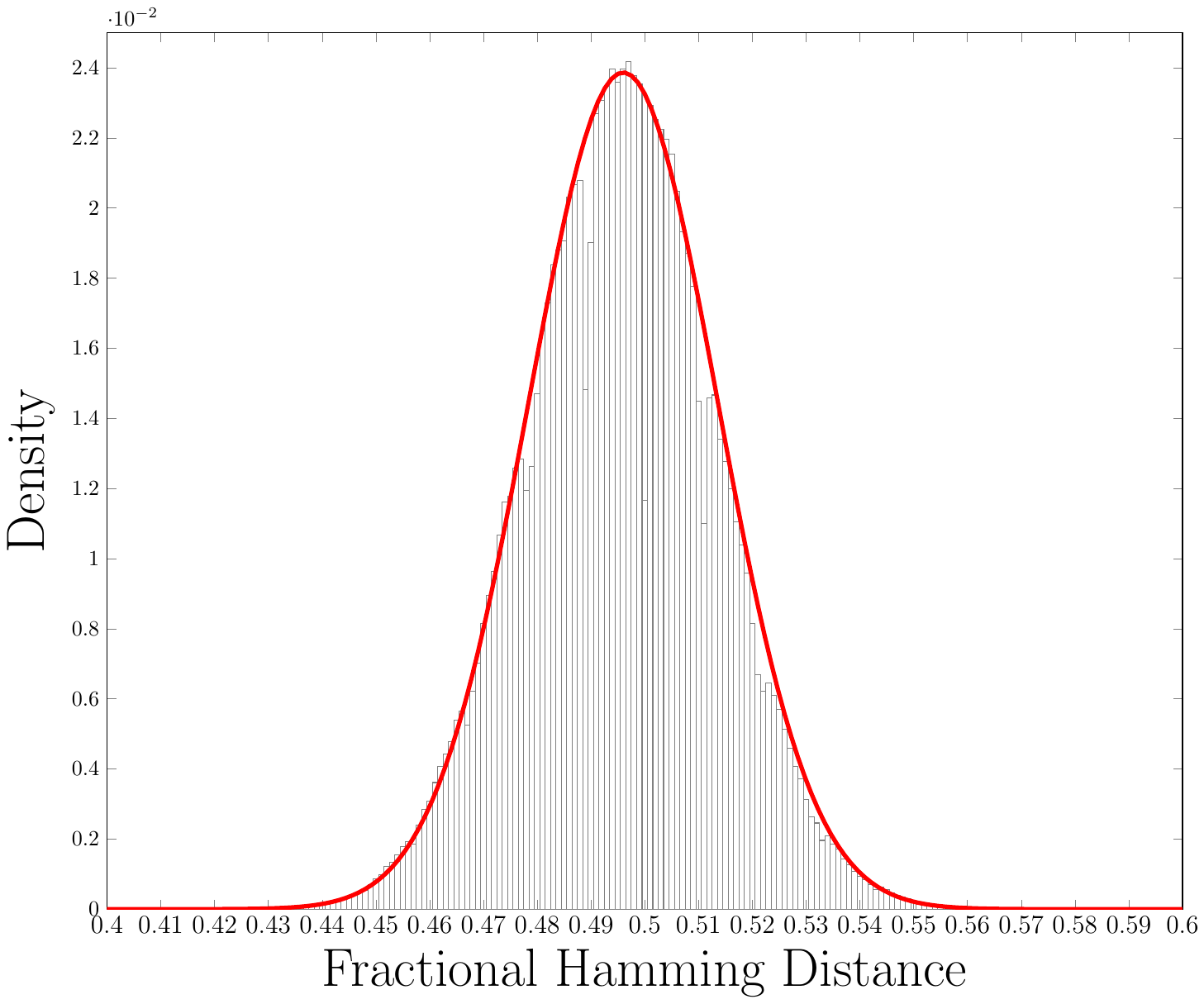}
	}
	\vskip\baselineskip
	
	\caption{\label{inter_intra_hamming} Histograms of Hamming distances in the benchmark dataset.}
\end{figure*}

Decidability for the dataset is $d' \approx 21$, which compares favourably to $d' \approx 14$ for iris recognition in the ideal condition~\cite{daughmanIRIS2}. 
The number of degrees of freedom is calculated based on Equation~\ref{eq:dof} as $N=807$, which means the entropy of the extracted fingerprints is 807 bits out of a total of 2048 bits. As compared to the 249 degrees of freedom for iris (which has the same size of 2048 bits), the fingerprint in our case is more unique and contains less redundancy. 
Figure~\ref{inter_intra_hamming} shows the histogram of same-group Hamming distance values on the left and the distribution of different-group Hamming distance values on the right. 
The diagram on the right also includes a binomial distribution curve with degrees of freedom $N=807$, mean $\mu = 0.495$, and standard deviation $\sigma = 0.018$. 
Evidently, the different-group distribution closely follows the binomial distribution. 

\textbf{PUF evaluations results. }
The PUF metrics results on the benchmark dataset are shown in Table~\ref{statistical_table} under the column labelled ``Benchmark Dataset''. 
It can be seen that in all metrics our dataset performed close to ideal values. 
For comparison, we also included in Table~\ref{statistical_table} the PUF metrics for two typical PUFs: Arbiter PUF, and Ring Oscillator PUF~\cite{statPUF}. 
This shows that our method provides fingerprints with good uniformity, randomness, reliability, steadiness, uniqueness, and bit-aliasing.

\textbf{Timing Results \& Usability.} 
Our paper fingerprinting method takes 1.30 seconds on average to prepare the photo, analyse the texture, and generate the fingerprint on a PC. This is reasonably fast. This is in contrast with the method in~\cite{clarksonfingerprint}, which requires four scans in different directions and then constructing a 3D surface model. 
Although the authors of~\cite{clarksonfingerprint} do not report timing measurements for their fingerprinting method, 3D modeling is generally considered a computationally expensive task~\cite{3Dmodeling}. 

The whole process of paper fingerprinting in our method is automatized and only requires a user to place the sheet of paper on the flat surface of the overhead projector and click a button to take a photo by a fixed camera. We note that this is only a proof-of-concept prototype to demonstrate the feasibility of extracting the fingerprint based on the textural patterns. One may improve the prototype in a practical application by tighter integration of various equipment components. For example, at a border control, when the official swipes a page in the passenger's passport through a slot, the slot may have the embedded light source on one side and a camera on the other side. When the page is in the slot, a unique fingerprint can be extracted. The fingerprinting area and orientation will be relatively fixed as it is determined by the dimensions of the slot. By comparing the extracted fingerprint with a reference sample (e.g., stored in the back-end system), the computer program can quickly determine if the passport page is genuine. In Section~\ref{security}, we will explain more on how to utilizing the unique paper fingerprint in authentication protocols.

\begin{table*}[]
	\small
	\centering
	\caption{False recognition rates of all datasets considering a fractional HD threshold of 0.4}
	\label{recognition_performance}
	\begin{tabular}{@{}l|c|c|cccccc@{}}
		\toprule
		Rate & Ideal & Benchmark & Rotated & Crumpled & Scribbled & Soaked & Heated & Mixed   \\ 
		&  Value & Dataset &  &  &  &  & & Light  \\ \midrule
		FAR                 & 0\% 	& 0\% 	& 0\% 		& 0\% 		& 0\% 	&0\% 	& 0\% 	& 0\% \\ 
		FRR                 & 0\% 	& 0\% 	& 0.32\% 	& 3.2\% 	& 0\% 	& 0\% 	& 0\% 	& 0\% \\
		
		\bottomrule
	\end{tabular}
\end{table*}

\begin{table*}[]
	\small
	\centering
	\caption{PUF metrics for all datasets and two typical PUFs}
	\label{statistical_table}
	\begin{tabular}{@{}l|c|cc|c@{}}
		\toprule
		{\footnotesize PUF} & {\footnotesize Ideal} & {\footnotesize Arbiter PUF (APUF)} & {\footnotesize Ring Oscillator PUF} & {\footnotesize Benchmark} \\ 
		{\footnotesize Metrics} & {\footnotesize Value} & {\footnotesize \cite{statPUF}} & {\footnotesize \cite{statPUF}} & {\footnotesize Dataset}\\ 
		\midrule
		{\footnotesize Average Uniformity}   & 0.5   & 0.556       & 0.505           & 0.466  \\
		{\footnotesize Average Randomness}   & 1.0   & 0.846       & 0.968           & 0.907   \\
		{\footnotesize Average Reliability}  & 1.0   & 0.997       & 0.991           & 0.945   \\
		{\footnotesize Average Steadiness}   & 1.0   & 0.984       & 0.985           & 0.938   \\
		{\footnotesize Average Uniqueness}   & 0.5   & 0.072       & 0.472           & 0.465   \\
		{\footnotesize Average Bit Aliasing} & 0.5   & 0.195       & 0.505           & 0.466   \\
		\bottomrule
	\end{tabular}
\end{table*}

\begin{table*}[]
	\centering
	\caption{Impact of Robustness Experiments on PUF metrics}
	\label{robustness_table}
	\scriptsize
	\begin{tabular}{@{}l|c|c|cccccc@{}}
		\toprule
		{\scriptsize PUF} & {\scriptsize Ideal} & {\scriptsize Benchmark} & {\scriptsize Rotated} & {\scriptsize Crumpled} & {\scriptsize Scribbled} & {\scriptsize Soaked} & {\scriptsize Heated} & {\scriptsize Mixed}  \\ 
		{\scriptsize Metrics} & {\scriptsize Value} & {\scriptsize Dataset} & & & &  &  & {\scriptsize Light} \\
		\midrule
		{\footnotesize Average Uniformity}   & 0.5   & 0.466   &  0.466  & 0.463    & 0.454     & 0.460  & 0.460 &  0.466   \\
		{\footnotesize Average Randomness}   & 1.0   & 0.907   &  0.906  & 0.896    & 0.873     & 0.877  & 0.890 &  0.907   \\
		{\footnotesize Average Reliability}  & 1.0   & 0.945   &  0.877  & 0.852    & 0.856     & 0.750  & 0.882 &  0.905   \\
		{\footnotesize Average Steadiness}   & 1.0   & 0.938   &  0.839  & 0.528    & 0.870     & 0.554  & 0.554 &  0.874   \\
		{\footnotesize Average Uniqueness}   & 0.5   & 0.465   &  0.465  & 0.470    & 0.468     & 0.463  & 0.461 &  0.465   \\
		{\footnotesize Average Bit Aliasing} & 0.5   & 0.466   &  0.466  & 0.463    & 0.454     & 0.460  & 0.460 &  0.466   \\
		\bottomrule
	\end{tabular}
\end{table*}

\section{Robustness Evaluations}
\label{robustness}
In this section we evaluate our method's robustness in non-ideal circumstances. First, we consider the robustness of our method against misalignment, i.e., in cases where the rectangular box is not aligned to the photo frame. Then, we consider the robustness of our method against paper being roughly handled in the following cases: the paper sheet is crumpled, some scribbling is done in the rectangular box, the sheet is soaked in water and dried afterwards, and the sheet is ironed after soaking and partially burnt. Finally, we consider the effect of using an alternative light source.
In the following, we give the details of each experiment and provide the biometric and PUF metrics in each of the cases. 

\subsection{Impact of Non-Ideal Data Collection}

\begin{figure*}
	\centering
	\subfigure[\label{envpic-normal} Benchmark]
	{\includegraphics[scale=0.124]{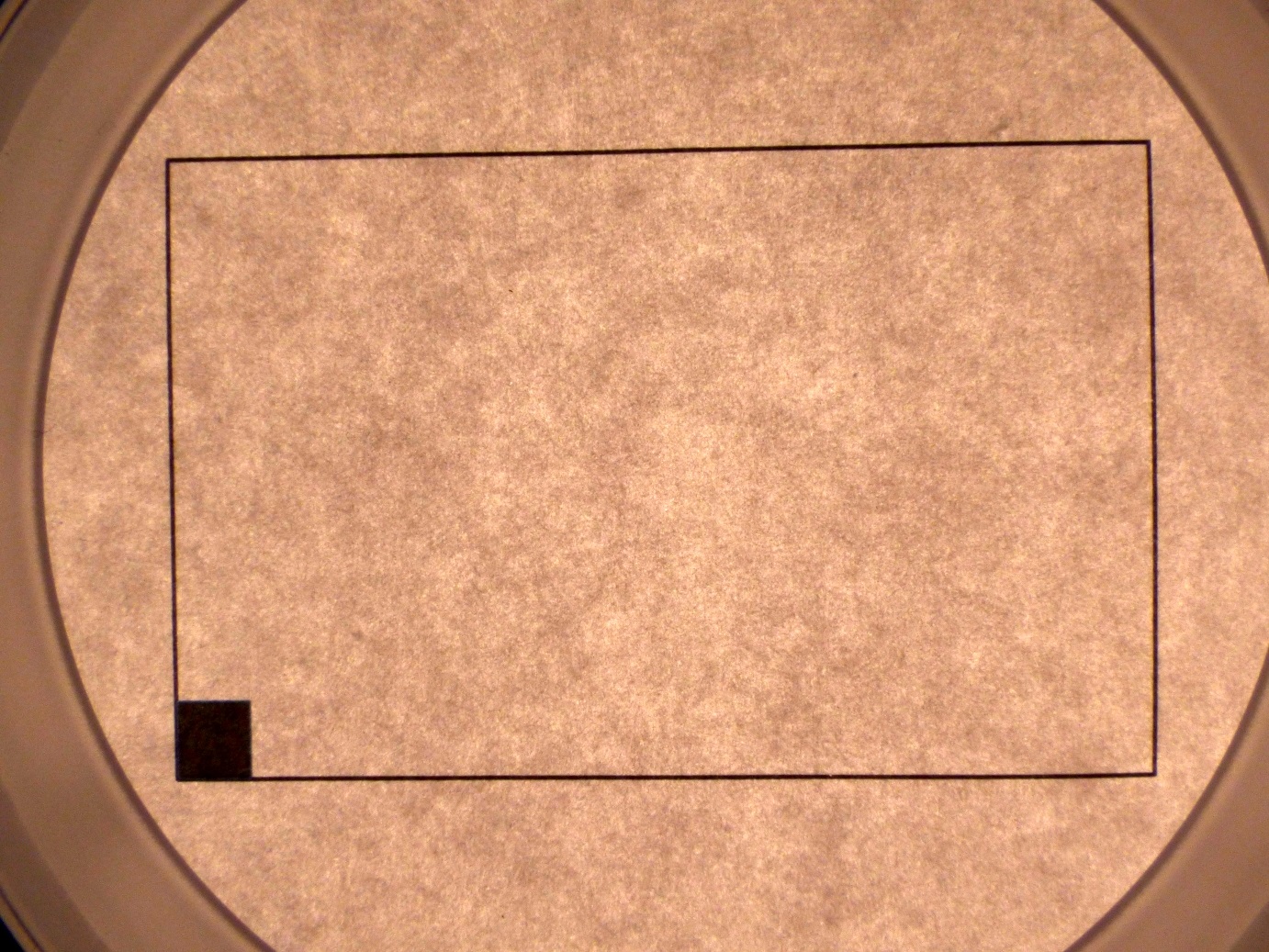}}
	\subfigure[\label{envpic-rotated} Rotated]
	{\includegraphics[scale=0.124]{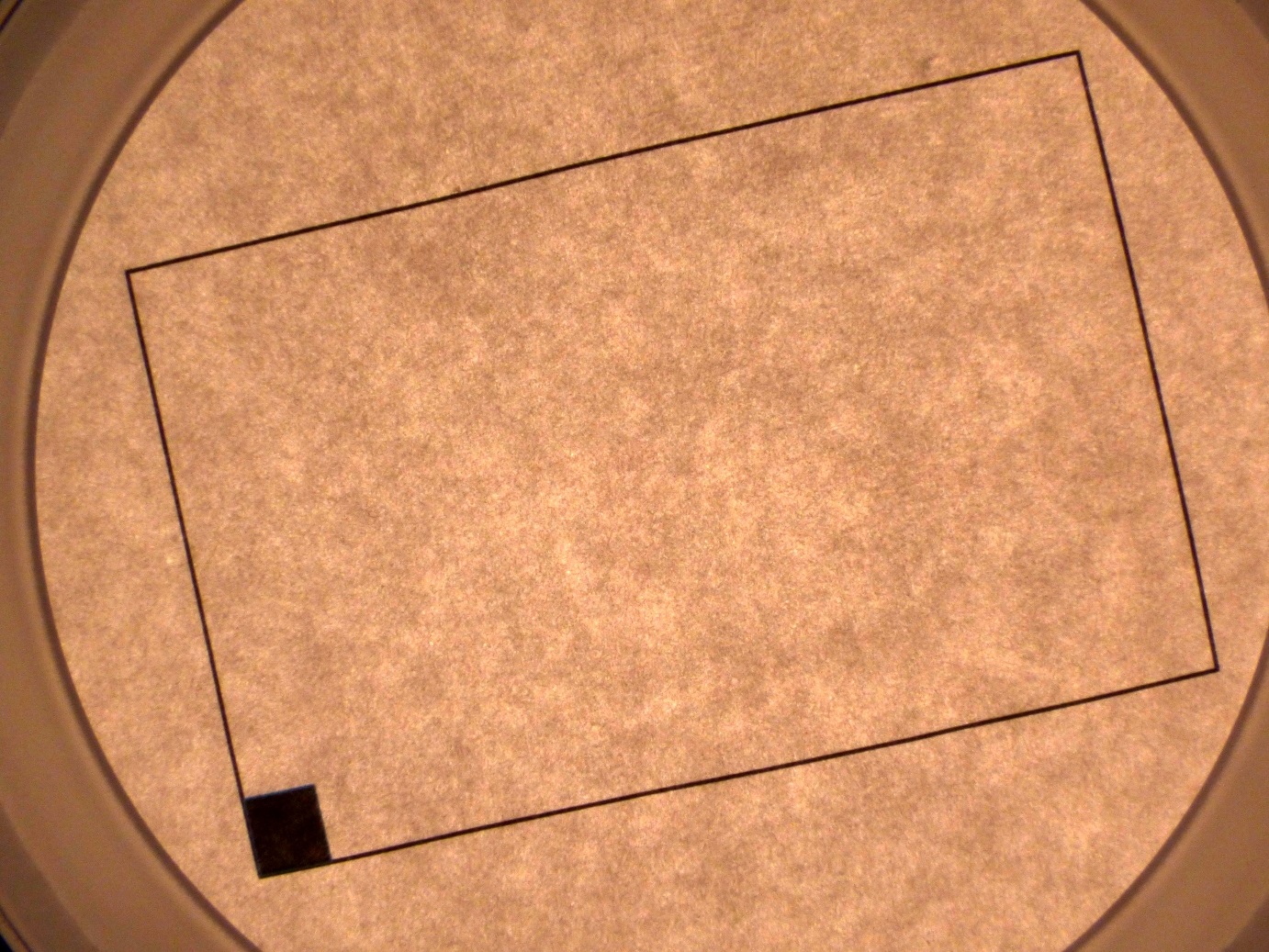}}
	\subfigure[\label{envpic-crumpled} Crumpled]
	{\includegraphics[scale=0.0375]{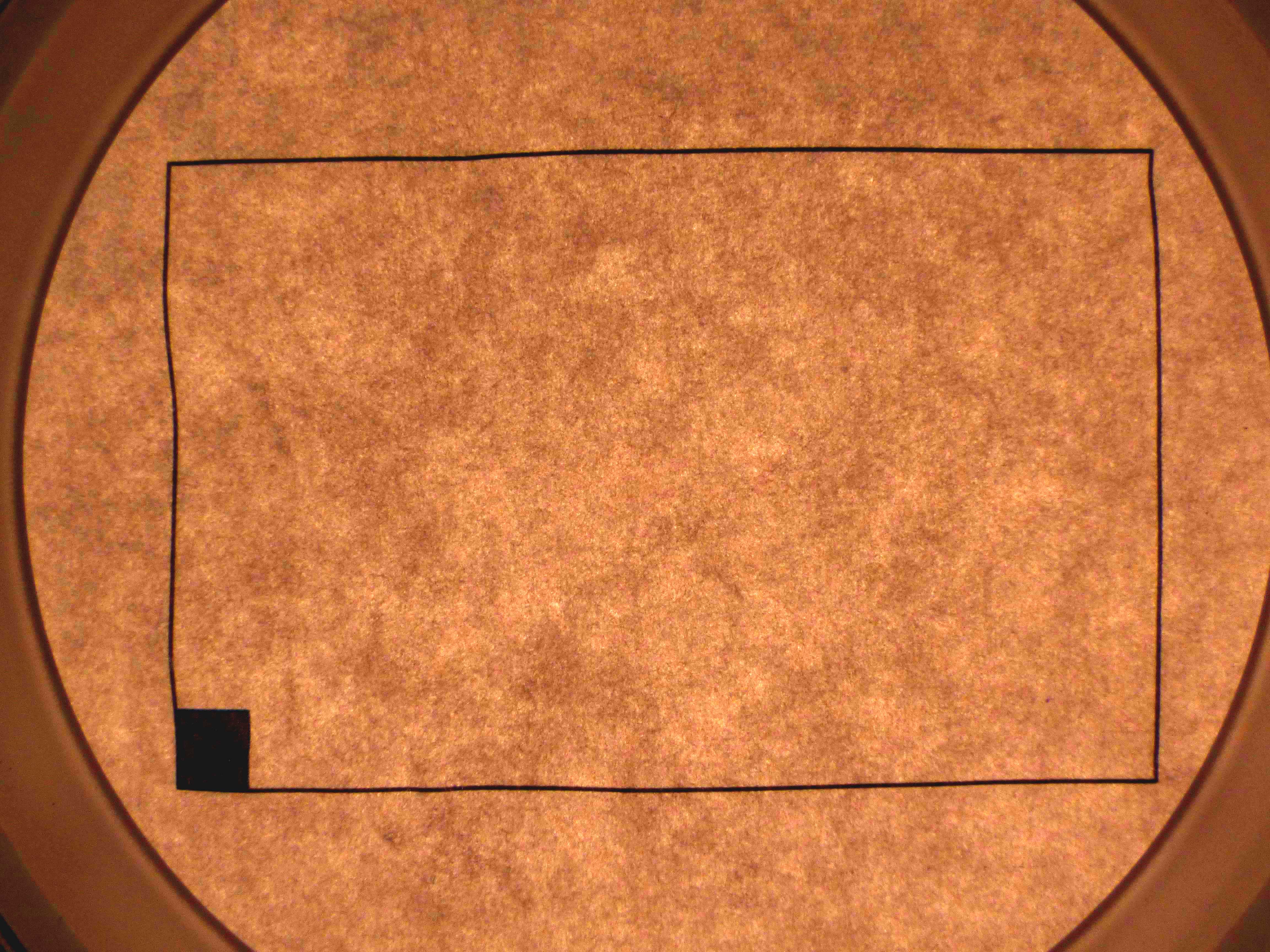}}
	\subfigure[\label{envpic-scribbled} Scribbled]
	{\includegraphics[scale=0.0375]{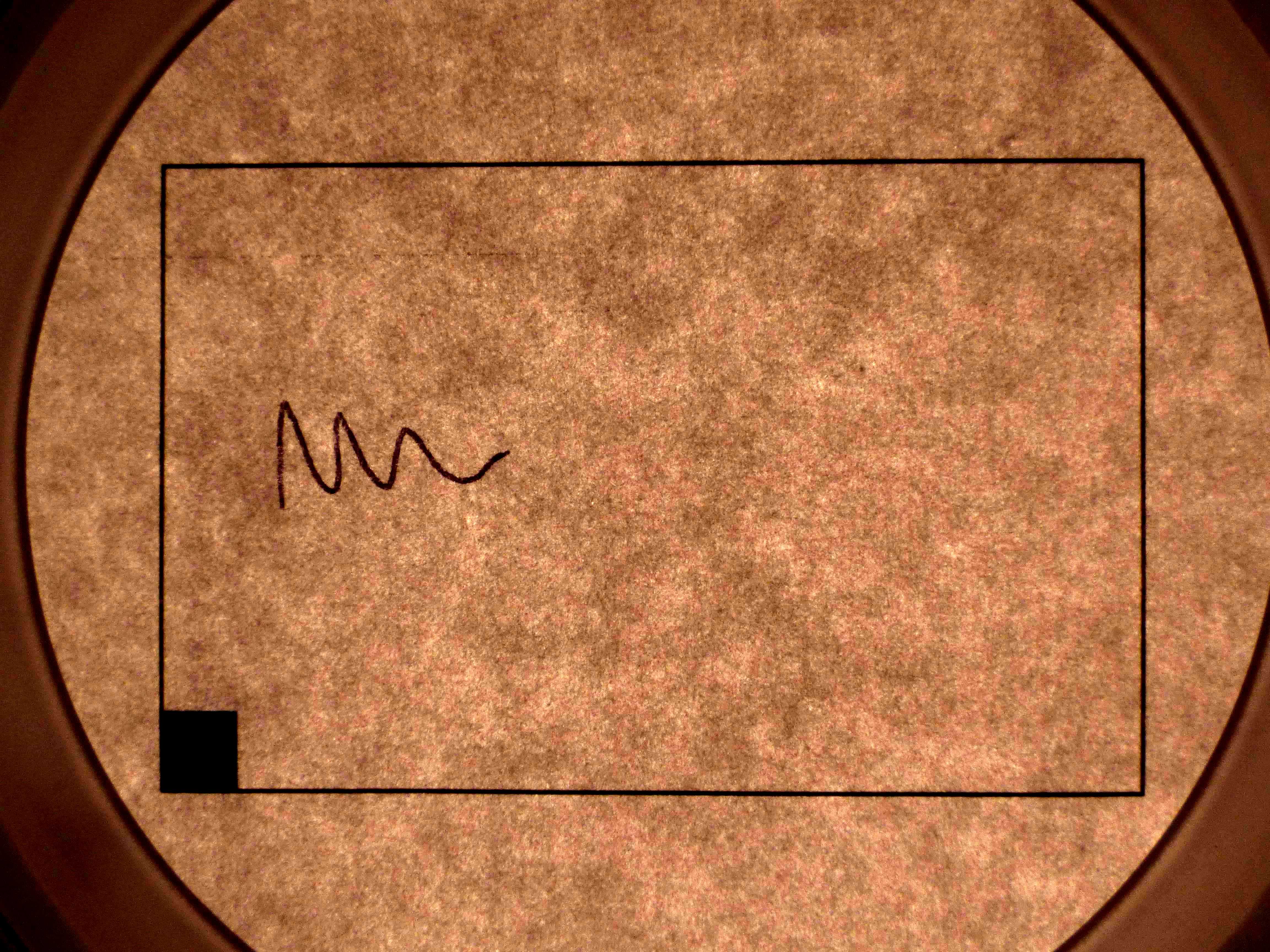}}
	\caption{The captured photo under near-ideal and non-ideal situations.}
	\label{envpic}
\end{figure*}

\textbf{Photo Rotation.} 
The orientation of the photo is the angle between the rectangular box and the photo frame. 
A rotated photo is shown in Figure~\ref{envpic-rotated}. 
The maximum rotation we can have such that the box is still fully captured within the boundary of the photo frame is around 12$^{\circ}$. 
We selected 10 paper sheets and collected 5 samples in each angle within 
$\{-12^{\circ},-11^{\circ},\ldots,0^{\circ},\ldots,,+11^{\circ},+12^{\circ}\}$. 
This gives us 125 samples per sheet, 1250 samples in total. 



Figure~\ref{ChangeRobustnessHamming} shows the Hamming distance distributions. 
As expected, the same-group and different-group distributions get slightly closer to each other in comparison with the benchmark dataset. 
However, decidability, although reduced, is still a healthy $d' \approx 8$. 
This shows that our image processing method is somewhat sensitive to the image rotation. 
We believe there is still room to improve the robustness against rotation, however
with the current method and based on a threshold of 0.4, the FAR is still 0\%, and the FRR is less than 1\%. 
These values can be found in Table~\ref{recognition_performance}. 


The PUF metrics are presented in Table~\ref{statistical_table}. 
The experiment dataset still has good uniformity, randomness, and bit-aliasing, but there is a slight drop in reliability, steadiness, and uniqueness compared to the benchmark dataset. 

The experiment shows that our method is robust against non-ideal data collection in terms of rotation. 
In comparison, Clarkson et al.\ do not report robustness against rotation and in fact require ``precise alignment of each surface point across all scans''~\cite{clarksonfingerprint}.

\subsection{Impact of Non-Ideal Paper Handling}
\begin{figure*}
	\centering
	\subfigure[Fitted distributions under rotation.]
	{
		\includegraphics[scale=0.22]{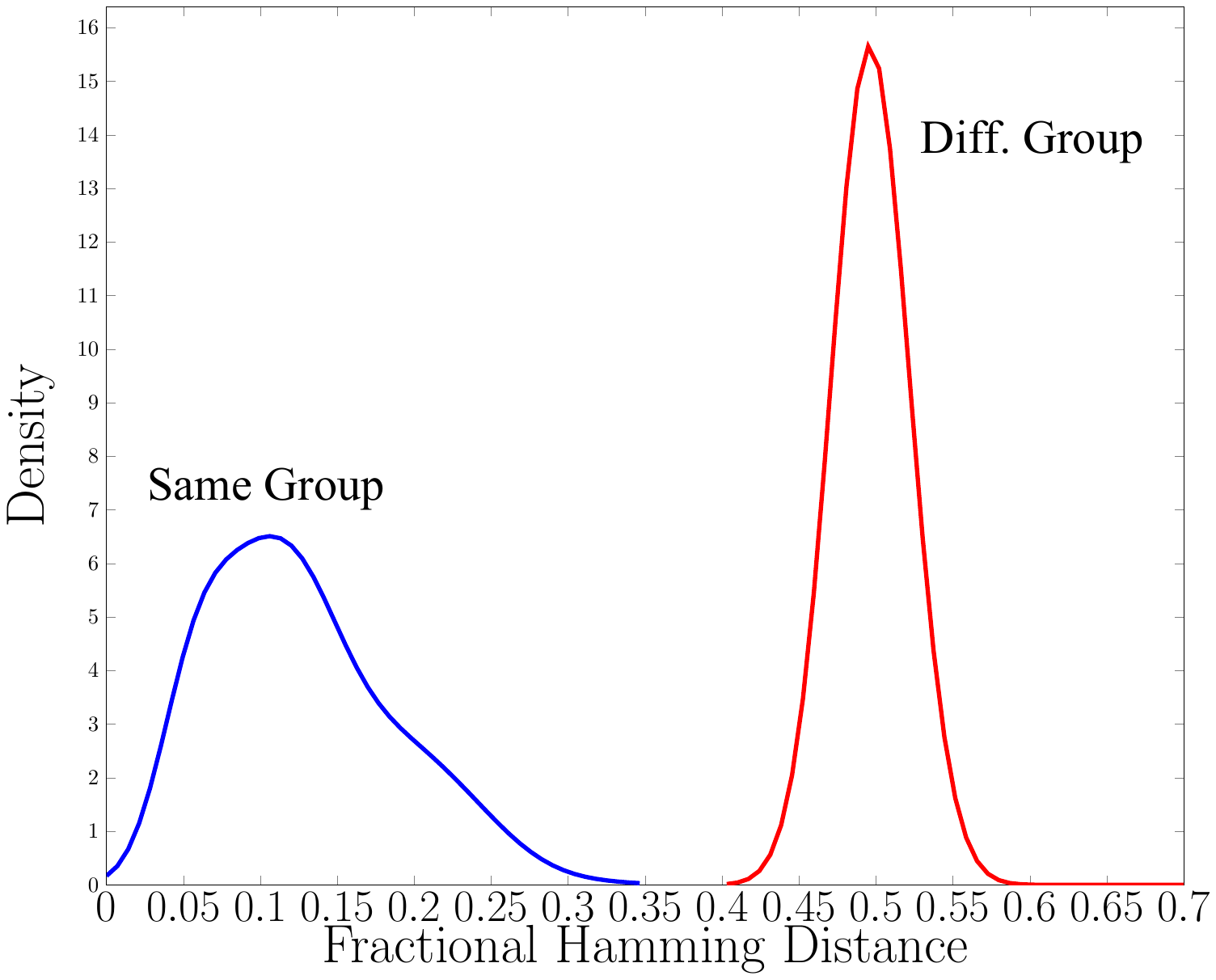}
	}
	\quad
	\subfigure[\label{ChangeRobustnessHammingTexture}Fitted distributions under non-ideal paper handling.]
	{
		\includegraphics[scale=0.22]{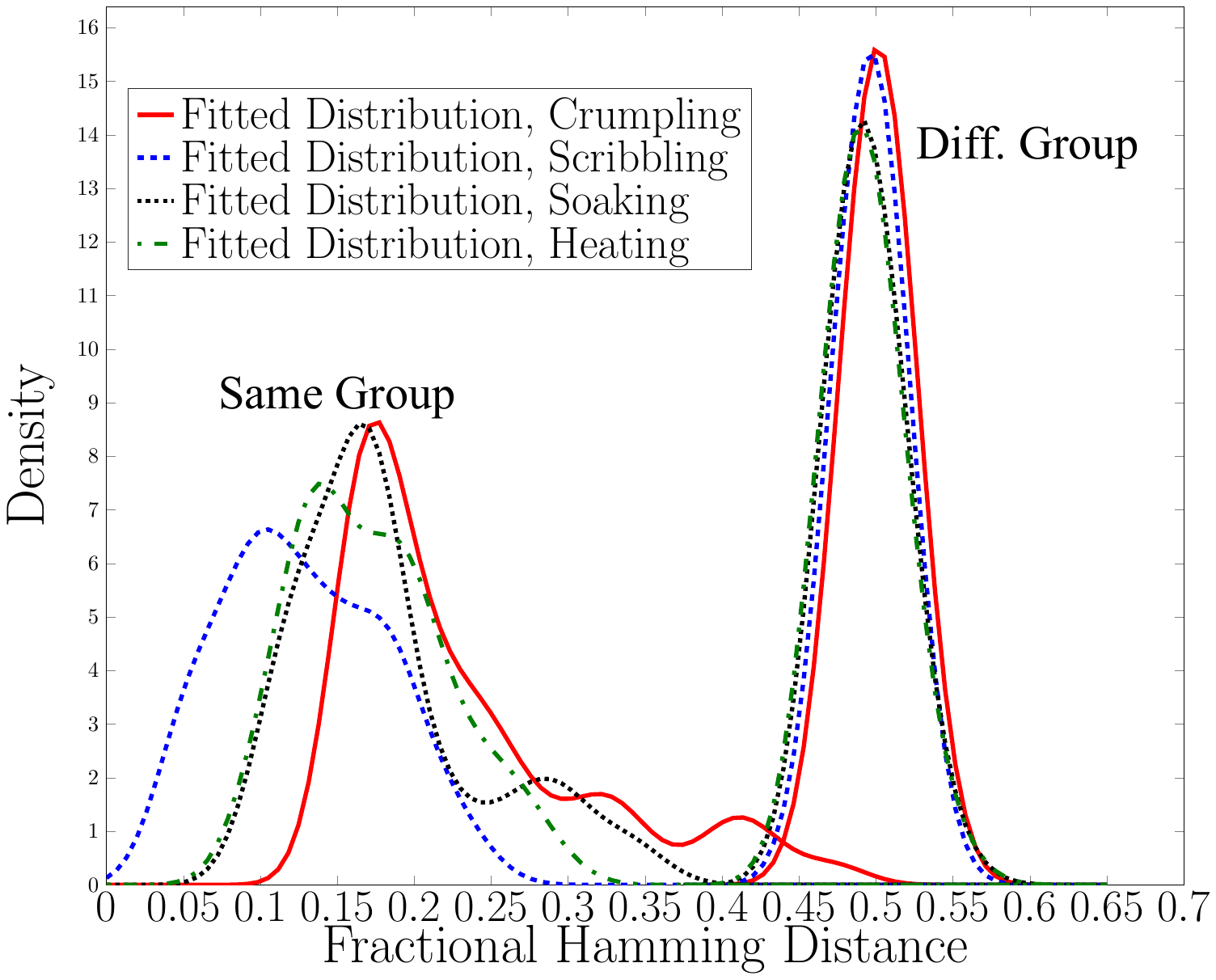}
	}
	\caption{\label{ChangeRobustnessHamming} The Hamming distance distributions for robustness experiments.}
\end{figure*}

In this section we investigate the robustness of our method against rough handling of paper sheet including crumpling, scribbling, soaking, and heating. 
For each of the experiments in this section, a set of 10 paper sheets are selected. 
For each paper sheet, 5 samples were taken before and 5 samples after the non-ideal handling of the paper sheet, adding up to a total of 100 samples per experiment. 
The same-group and different-group distributions under the test conditions of crumpling, scribbling, soaking and heating are shown in Figure \ref{ChangeRobustnessHamming}.
For readability, we opt to show fitted curves for the distributions. 
These curves are non-parametric fits with a threshold bandwidth of 0.02 (i.e., the distributions are merely smoothed).

\textbf{Crumpling.}
In this experiment, we crumpled our paper sheets to the extent that the borders of the rectangular box were visibly distorted. 
We did not try to smooth out the sheet surface after crumpling. 
An example of a photo taken from a crumpled paper sheet can be seen in Figure~\ref{envpic-crumpled}. 

The resulting Hamming distance distributions are shown in Figure~\ref{ChangeRobustnessHamming}. 
Decidability is $d' \approx 4.6$. 
Based on the threshold of 0.4, the FAR is still 0\%, and the FRR is 3.2\%. 
These values can be found in Table~\ref{recognition_performance}. 

The PUF metrics are presented in Table~\ref{robustness_table}. 
The experiment dataset still has good uniformity, randomness, and bit-aliasing, but there is a slight drop in reliability and uniqueness and a bigger drop in steadiness compared to the benchmark dataset. 


\textbf{Scribbling.}
In this experiment, we drew random patterns with a black pen over all samples such that each pattern covers around 5\% of the box area. 
An example of such scribbling can be seen in Figure~\ref{envpic-scribbled}. 
Our preprocessing phase successfully identifies the scribbled area in the mask in all samples. 

The resulting Hamming distance distributions are shown in Figure~\ref{ChangeRobustnessHamming}. 
The maximum same-group Hamming distance is 0.25 and the minimum different-group Hamming distance is 0.45. 
The distributions are well-separated. 
Decidability is $d' \approx 9.7$. 
Based on the threshold of 0.4, the FAR is still 0\%, and the FRR is also 0\%. 
These values can be found in Table~\ref{recognition_performance}. 

The PUF metrics are presented in Table~\ref{robustness_table}. 
The experiment dataset still has good uniformity, randomness, and bit-aliasing, but there is a slight drop in reliability, steadiness, and uniqueness compared to the benchmark dataset. 



\textbf{Soaking.}
In this experiment, we submerged the paper sheets in tap water for around 20 seconds. 
Then, we let them dry naturally 
and collected the after-soaking samples from the dried sheets. 

The resulting Hamming distance distributions are shown in Figure~\ref{ChangeRobustnessHamming}. 
The maximum same-group Hamming distance is 0.36 and the minimum different-group Hamming distance is 0.44. 
The distributions are well-separated. 
Decidability is $d' \approx 6.8$. 
Based on the threshold of 0.4, the FAR is still 0\%, and the FRR is also 0\%. 
These values can be found in Table~\ref{recognition_performance}. 

The PUF metrics are presented in Table~\ref{robustness_table}. 
The experiment dataset still has good uniformity, randomness, and bit-aliasing, but there is a slight drop in reliability and uniqueness and a bigger drop in steadiness compared to the benchmark dataset. 


\textbf{Heating.}
In this experiment, we ironed all the papers from the soaking experiment for at least 20 seconds, to the extent that in some cases there was a clearly visible colour change (to light brown) and the paper was partly burnt. 

The resulting Hamming distance distributions are shown in Figure~\ref{ChangeRobustnessHamming}. 
The maximum same-group Hamming distance is 0.30 and the minimum different-group Hamming distance is 0.44. 
The distributions are well-separated. 
Decidability is $d' \approx 8.6$. 
Based on the threshold of 0.4, the FAR is still 0\%, and the FRR is also 0\%. 
These values can be found in Table~\ref{recognition_performance}. 

The PUF metrics are presented in Table~\ref{robustness_table}. 
The experiment dataset still has good uniformity, randomness, and bit-aliasing, but there is a slight drop in reliability and uniqueness and a bigger drop in steadiness compared to the benchmark dataset. 


\textbf{Summary.}
Taking all the above results into consideration, we can see that our method shows the strongest robustness against scribbling. 
Both the biometric and PUF measures support this observation. 
The Hamming distance distributions are well-separated and all PUF metrics remain close to ideal values. 
Fingerprinting is also fairly robust against rotation, soaking, and heating. 
There is no or negligible false rejection rates and all PUF metrics possibly except for steadiness remain close to ideal values. 
Crumpling seems to pose the strongest challenge to robustness. 
Although false rejection rate is 3.2\% and steadiness is not ideal, the method is still able to provide 0\% false acceptance rate and healthy PUF metrics otherwise. 

Focusing on biometric metrics, authentication rates remain perfect or nearly perfect under all robustness tests. 
This means our method provides a promising candidate for paper-based document authentication in practice which is able to cope with non-ideal sample collection and rough handling. 

Focusing on PUF metrics, space and device dimension metrics stay close to ideal values under all tests, which indicates that the quality of fingerprint bits are still good and the sheets remain clearly distinguishable from one another. 
Time dimension metrics remain close to ideal values for rotation and scribbling, but steadiness and in some cases reliability drops as a result of crumpling, soaking, or heating. 
This is expected as crumpling, soaking, and heating physically change the paper sheets.

\subsection{Impact of a Different Light Source}
\label{lightBoxSource}

The light source should be bright enough to reveal the texture patterns in a paper sheet. In the proof-of-concept experiments, we used an overhead projector, however, the equipment is relatively bulky and expensive. Questions remain if there are cheaper ways to obtain the light source and if the results are robust against using a different light source. To investigate this, we purchased a commodity light box (tracing pad) from Amazon for \pounds 49.99 (see Figure~\ref{lightbox}). Then, we used the same paper sheets as in the benchmark dataset--excluding 10 paper sheets that were used in other robustness tests--to collect a new set of samples using the new light source. We followed the same data collection procedure as before. 

Due to the difference in the light intensity, the camera setting needs to be adjusted. In particular, we altered the exposure time to 1/500 seconds and F-stop to f/5. These values were automatically recommended by the camera, so we simply accepted them. The exposure time is the duration that the shutter takes to capture a photo and F-Stop is the radius of the lens diaphragm; both of them are inspired by the way human eyes react to a light source. These modifications in the camera setting were necessary because of the change in the intensity of the light source. The final dataset included 900 captured images, 10 samples from each paper sheet.

\begin{figure*}
	\centering
	
	\subfigure[Hamming distance distributions for the light box dataset, plus the binomial curve with $N=846$, $\mu = 0.496$ , $\sigma = 0.017$%
	]
	{
		\includegraphics[scale=0.22]{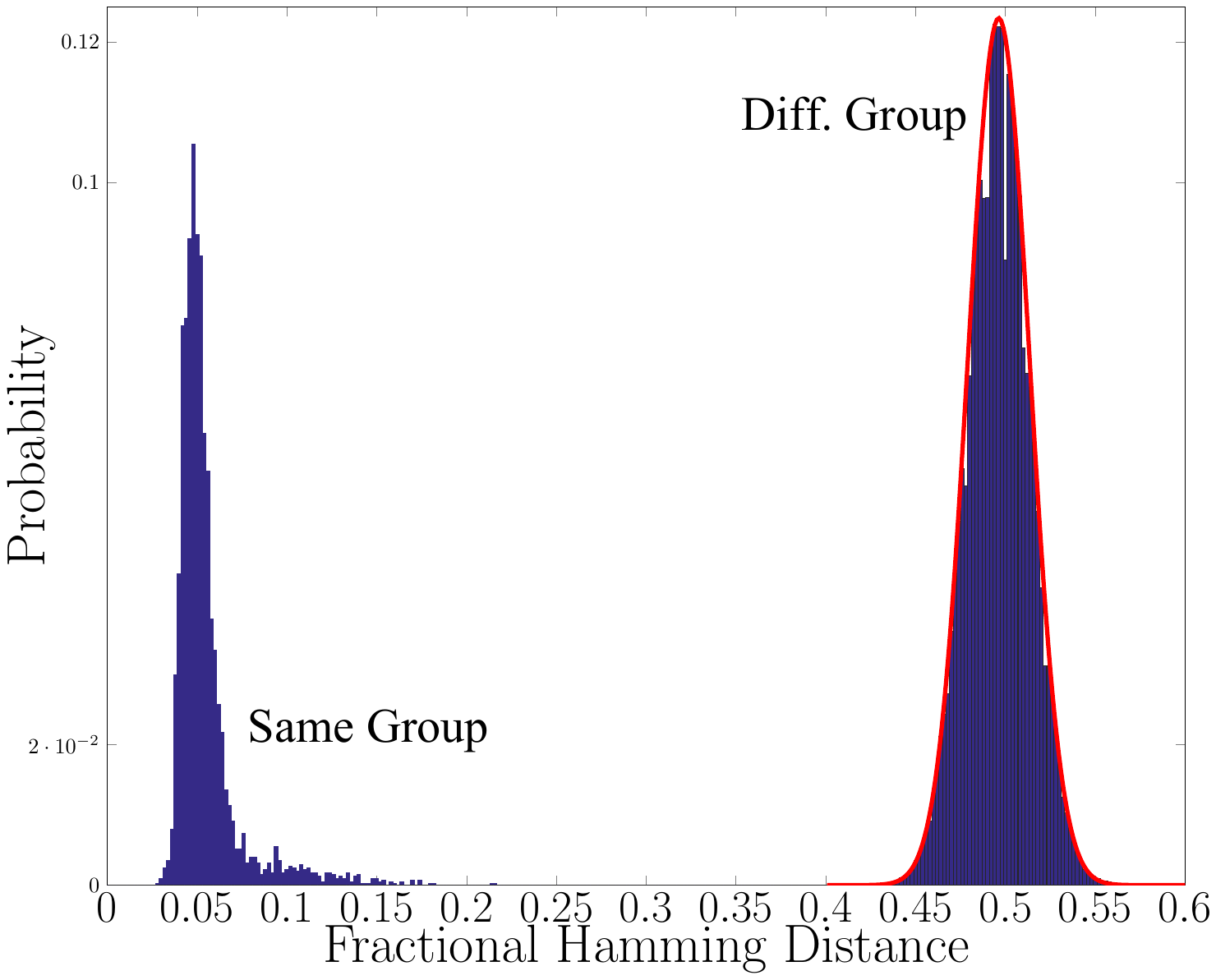}
		\label{LightSourceHamming}
	}
	\quad
	\subfigure[Hamming distance distributions for the mixed light box and projector dataset, plus the binomial curve with $N=836$, $\mu = 0.496$ , $\sigma = 0.017$]
	{
		\includegraphics[scale=0.22]{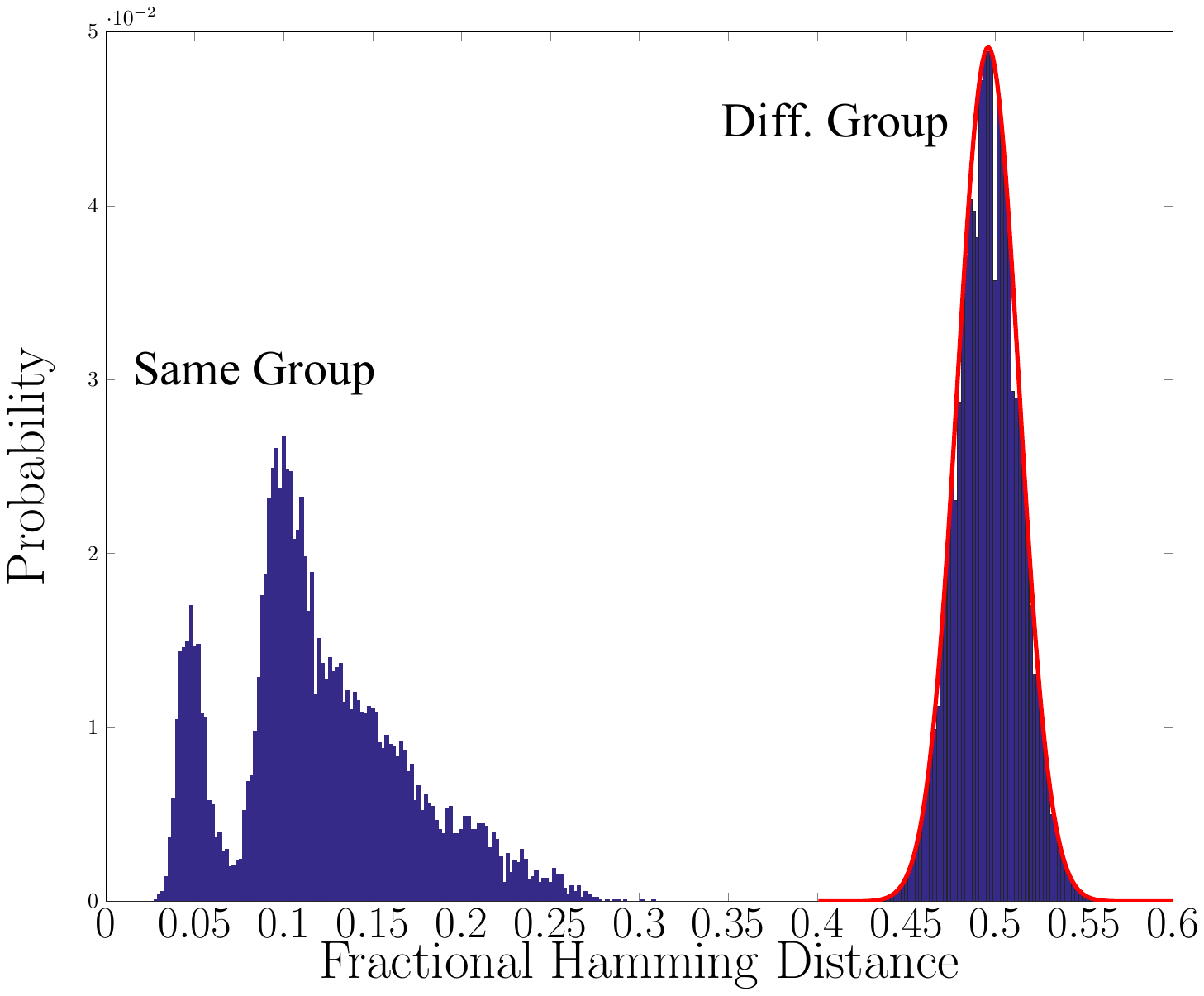}
		\label{LightSourceandProjector}
	}
	\vskip\baselineskip
	
	\caption{\label{lightbox_inter_intra_hamming} Distributions of HDs for the light box experiment.}
\end{figure*}

Figure~\ref{LightSourceHamming} shows the Hamming distance distributions using the light box. The same-group and different-group distributions are well-separated from each other. Applying the biometric metrics, our analysis shows the decidability $d' \approx 24$ and the number of the degrees of freedom $DoF \approx 846$, both slightly higher than those obtained with the overhead projector. Based on the threshold of 0.4, the FAR and FRR are still 0\%. These values can be found in Table~\ref{recognition_performance}. 

The PUF metrics are presented in Table~\ref{robustness_table}. The new experiment results show that all PUF metrics are comparable to those obtained earlier in the benchmark dataset. 

Figure~\ref{LightSourceandProjector} shows the Hamming distance distribution by combining the light box and overhead projector datasets. The number of the degrees of freedom is roughly unchanged at $DoF \approx 836$. However, the same-group data become noisier because of mixing two different light sources. The decidability drops to 10. Despite of the mix of different light sources, the same-group and different-group histograms are still clearly separated. The maximum Hamming distance for the same-group samples is 0.31 while the minimum Hamming distance of the different-group is 0.42.

The experiment shows that our method is robust against different light sources, as long as the camera settings are set correctly. 

\section{Authentication protocols}
\label{security}
In this section, we explain authentication protocols based on the extracted paper fingerprint, and discuss their practical performance.

\subsection{Trust assumptions}

Our fingerprinting technique may be applied in a range of applications, e.g., to prevent counterfeiting of paper currency, passports, certificates, contracts and official receipts. The secure use of the fingerprint is based on two assumptions. Both assumptions are generally required in biometrics and physical unclonable functions (PUF) applications.

The first assumption is physical ``unclonability''. We assume it is infeasible to physically clone a paper sheet with the same paper texture. The paper texture is formed from randomly interleaved wooden particles, as a naturally occurring outcome of the paper manufacturing process. This process can not be precisely controlled. Repeating exactly the same process to produce the same paper texture is considered to be prohibitively expensive, if not impossible~\cite{pufterm1}. 

The second assumption is about a trustworthy measuring process. Take the human fingerprint authentication as an example. If an attacker is able to deceive the scanner by presenting a gummy finger, the security guarantee based on the ``unclonability'' assumption will be lost. In any biometric or PUF application, it is important to ensure that the measurement is performed on a real object and  a \emph{fresh} measurement is acquired. In practice, this is often realized through the human supervision in the process or by using specialized equipment (e.g., iris scanners with embedded liveness test). In the case of paper documents, visual inspection can be applied to check that they are made of paper and the paper fiber texture has not been tampered with. An attacker may try to interfere with the texture measurement by printing patterns on the paper surface. Using today's commodity printers, it seems unlikely that an attacker is able to print patterns that are precise at the pixel level under the microscopic view of a high-resolution camera (since the print head cannot be precisely controlled and each printed dot tends to be in a scattered pattern due to imperfection of the printing process; see~\cite{clarksonfingerprint}). However, when the measurement is not guaranteed to be coming from real paper texture, the acquisition process is no longer trustworthy -- an attacker can at least deny the authentication by printing random patterns with strong contrast on the paper. This threat can be addressed by checking that the intended area for authentication is free from overprinting. 


\subsection{Comparison based on Hamming distance}
\label{comparisonHamming}
A straightforward application of authenticating a paper fingerprint is based on comparing the Hamming distance between two fingerprints. It consists of two phases. In the first phase, a paper fingerprint, along with a mask, is extracted from the textural patterns as the template and stored in a  database. In the second phase, given a provided paper sheet, the same fingerprinting algorithm is followed to output a new fingerprint and a mask. Depending on the applications, there are two types of authentication modes: verification or recognition. 

Verification works on a one-to-one comparison. This assumes the reference to the stored template is known (as it is often provided by the authenticating subject). Hence, once the template is retrieved, it is a straightforward comparison between two fingerprints based on their Hamming distance as explained in Equation~\ref{maskedHamming}. This comparison determines if the presented paper sheet is the same as the one registered earlier.

By contrast, recognition works on a one-to-many comparison. In this case, the reference to the pre-stored template is unknown. Hence, the program searches throughout the database, comparing the extracted fingerprint exhaustively with each of the stored templates in order identify a match where the Hamming distance is sufficiently small. This is the same as how iris recognition works. 

In terms of accuracy, the recognition mode is far more demanding than the verification mode, because the false accept rate accumulates with the size of the database. As an illustration, let $P_1$ be the false acceptance rate for one-to-one matching in the verification mode. Assume $P_1$ is very small. Let $P_n$ be the false acceptance rate in the recognition mode for a database of $n$ records. 
\begin{eqnarray*}
	P_n & = & 1 - (1-P_1)^n \\
	& \approx & n\cdot P_1
\end{eqnarray*}
The above equation shows that the accumulative false acceptance rate in the one-to-many mode increases roughly linearly with the size of the database~\cite{daughmanIRIS2}. Hence, for the one-to-many matching to work accurately, the false acceptance rate for the one-to-one comparison must be extremely small.

For the paper fingerprints extracted in our proposal, they have sufficient entropy to support precise recognition even for an extremely large database. Based on the binomial distributions with 807 degrees of freedom, the false acceptance rates for comparing two paper fingerprints are listed in Table \ref{falseaccept}. If we opt to maintain $P_n < 10^{-6}$ for the recognition mode as stated in \cite{daughmanIRIS2}, our algorithm can easily support searching a database of 3 quintillions ($3 \times 10^{18}$) fingerprints at a threshold of $0.32$. By comparison, for the same accuracy ($<10^{-6}$) and the same threshold ($0.32$), iris recognition can only support a database of only 26 iris codes. (As stated in \cite{daughmanIRIS2}, for a database of a million iris codes, the threshold needs to be adjusted downwards to below $0.27$ to keep the false accept rate under $10^{-6}$). Because of the much higher degrees of freedom of paper fingerprints, they can be used for the recognition application at a much larger scale than the iris biometric.

\begin{table}[t]
	
	\caption{\label{falseaccept}False Acceptance Rate (FAR) for comparing two fingerprints}
	\centering
	\begin{tabular}{c c}
		\toprule
		HD Threshold & False acceptance rate \\
		\hline
		0.30  & $7.1 \times 10^{-31}$ \\
		0.31  & $5.3 \times 10^{-28}$ \\
		0.32    & $2.7 \times 10^{-25}$ \\
		0.33  & $1.0 \times 10^{-22}$ \\
		0.34    & $2.5 \times 10^{-20}$ \\
		0.35  & $4.5 \times 10^{-18}$ \\ 
		0.36    & $5.8 \times 10^{-16}$ \\
		0.37  & $5.2 \times 10^{-14}$ \\ 
		0.38    & $3.3 \times 10^{-12}$ \\
		0.39  & $1.5 \times 10^{-10}$ \\
		0.40    & $5.2 \times 10^{-9}$ \\
		\bottomrule 
	\end{tabular}
\end{table}

\subsection{Paper fingerprint encryption}

One limitation with the previous verification/recognition method is that the template is stored in plaintext in the database. When the plaintext template is revealed, it may cause degradation of security. This is especially the case with biometrics, since biometric data is considered private to each individual. Paper fingerprints are essentially ``biometrics'' of paper. One established technique in biometrics is through \emph{biometric encryption}. Similarly, we can apply the similar technique to realize \emph{fingerprint encryption}. We will present one concrete construction and show that because paper fingerprints have much higher entropy than even the most accurate biometric in use (iris), the corresponding encryption scheme is able to provide much higher security assurance as well. 

\begin{figure}
	\centering
	\includegraphics[scale=0.33]{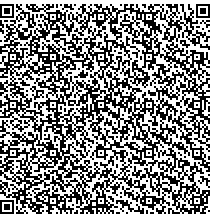}
	\caption{\label{qr} Generated QR Code in the authentication protocol. This QR code contains the encrypted fingerprint, $H(k)$ and a digital signature for both items.}
\end{figure}

Our construction is based on Hao et al.'s scheme \cite{fengBio}. This work is inspired by Juels et al.~\cite{JuelsECCAuthentication} and has been successfully implemented in iris recognition. It comprises two phases. In phase one, the program extracts a paper fingerprint from the paper texture as a reference $f_a$. It then generates a random key $k$ (140 bits), and expands the key to a pseudo fingerprint $f_p = \mathrm{ErrorCC}(k)$ (a 2048-bit codeword) where $\mathrm{ErrorCC}$ is an error-correction encoding scheme based on Hadamard-Reed-Solomon. Our analysis shows there is a combination of block and random errors in our fingerprints; therefore, we selected a concatenated approach. The choice of 140 bits $k$ is a balance between security (minimum 128 bit security for the secret key) and performance, as well as considering the special parametric requirements for a concatenated code scheme to work at a desired level of error correction. Subsequently, the scheme computes an encrypted fingerprint $r = f_a \oplus f_p$. In addition, the program computes $h = H(k)$ where $H$ is a secure one-way hash function. Finally, the program stores $r$ and $h$ in the database. Alternatively, $r$ and $h$ can be stored in a 2-D barcode printed on paper. The advantage of doing so is to allow authentication in the off-line mode. In this case, an additional digital signature $s$ should be included to prove the authenticity of data in the barcode. At this stage, the original template $f_a$ and the random key $k$ can be safely deleted. The registration process is summarized in Algorithm~\ref{registration}. In Figure~\ref{qr}, we show a QR code generated from the registration phase in our prototype implementation.

The second phase is authentication. In this phase, data from the 2-D barcode is first read and the digital signature verified. A paper fingerprint $f_s$ is extracted from the provided paper sheet. The program then computes:

\begin{eqnarray*}
	f_s \oplus r &=& f_s \oplus (f_a \oplus  \mathrm{ErrorCC}(k)) \\
	&=& (f_s \oplus f_a) \oplus \mathrm{ErrorCC}(k) \\
	&=& e \oplus \mathrm{ErrorCC}(k)
\end{eqnarray*}

In the above equation, $e$ can be regarded as ``noise'' added to the codeword $\mathrm{ErrorCC}(k)$. As we explained earlier, the Hamming distances between same-paper fingerprints typically range from $0$ to $0.25$. In the definition of the Hadamard-Reed-Solomon code, we follow the same coding parameters as in~\cite{fengBio}. The resultant error correction code is capable of correcting up to 27\% error bits in a 2048-bit codeword. Hence, by running the Hadamard-Reed-Solomon decoding scheme, the error vector $e$ can be effectively removed, and the original $k$ can be recovered error-free. The correctness of the decoding process can be verified by comparing the obtained $k$ against the retrieved $H(k)$. This authentication process is summarized in Algorithm~\ref{verification}.

\begin{algorithm}
	\caption{\label{registration} Registration}
	
	Generate Random key $k$ \;
	Generate Reference Paper Fingerprint $f_a$\;
	Expand key $k$ to Pseudo Fingerprint $f_p$  \;
	Calculate $r = f_a \oplus f_p$ \;
	Calculate $h = H(k)$ \;
	Calculate Digital Signature $s = Sig(r, h)$ \;
	Store $(r,h,s)$ in a 2-D barcode \;
	
\end{algorithm}


\begin{algorithm}
	\caption{\label{verification} Verification}
	Read $r$, $h=H(k)$ and $s=Sig(r,h)$ \;
	\eIf {Signature Verification Success}{
		Generate Paper Fingerprint $f_s$ \;
		Calculate $f'=f_s \oplus r$ \;
		Acquire $k'$ by decoding $f'$ \;
		Calculate H($k'$) \;
		\eIf {H($k'$)==H(k)}{
			 Success \;
		}{
		 Failure \;
	}
}{
 Failure \;
}
\end{algorithm}

The key feature of the above ``fingerprint encryption'' scheme is that it preserves the secrecy of the fingerprint template since it forms the basis for authentication. In this way, no fingerprint template is stored in the plain form. As an example for comparison, without using this encryption scheme, the barcode would contain the plain fingerprint template. Once in the line of sight to an attacker, the barcode can be trivially read say by using a video camera, hence the template will be stolen. With the encryption scheme applied, the attacker would need physical access to the paper in order to take a close-up snapshot of the fiber textures with a bright light source shining underneath the paper. This makes the attack significantly more difficult to carry out in practice without the user noticing it.


Hence, the application of privacy preserving protocol for authentication avoids storing the texture structure in the plain text form. The goal here is to protect the paper texture from an attacker who does not have physical access to the paper sheet itself. 
An adversary who has access to the barcode printed on the paper can read all data including an encrypted fingerprint $r = f_a \oplus \mathrm{ErrorCC}(k)$. One potential problem as highlighted in \cite{fengBio} is that if the fingerprint $f_a$ contains significant correlations between bits, $r$ may leak information about the fingerprint. The authors of \cite{fengBio} use the iris code as an example to illustrate that due to a high level of redundancy in iris codes, the encrypted iris code only has a lower-bound security of 44 bits. However, 44 bits security is not sufficient to satisfy high security requirements. As a result, the encrypted iris code (also called the secure sketch in the PUF literature) should not be published as public data; instead, it should be stored in a personal token as suggested in \cite{fengBio}.

The above limitation with the iris codes does not apply in our case. Although the paper fingerprint defined in our work has the same size as an iris code (2048 bits), it has much higher degrees of freedom (807 as compared to 249). Following the same sphere-packing bound as defined in \cite{fengBio}, we estimate the lower-bound security for the encrypted fingerprints as follows. Here, the lower-bound security refers to the minimum efforts required for a successful brute-force attack, under the assumption that the attacker has perfect knowledge of the correlations within the document paper sheet's fingerprint, hence the uncertainty (or entropy) about the fingerprint is 807 bits instead 2048 bits. 
The error correction capability for the Hadamard-Reed-Solomon code allows correcting up to 27\% error bits. So in principle the attacker only needs to guess a fingerprint that is within the Hamming distance of $807\times 0.27 \approx 218$ bits to the correct fingerprint. Following the estimation method in~\cite{fengBio}, based on the sphere-packing bound \cite{sphereBoundingHamming}, the minimum guess effort with $z=807$ and $w=218$ is calculated with the following equation:

\begin{equation}
\mathcal{G} \geq \frac{2^z}{\sum_{i=0}^{w}
	\begin{pmatrix}
	z\\ i
	\end{pmatrix}}
=  2^{133}
\end{equation}

The above bound states that an attacker with full knowledge about fingerprint correlations and the error correction process would need at least $2^{133}$ attempts in order to uncover the original fingerprint used in the registration and the random key $k$. This 133-bit security is much higher than the 44-bit security reported in \cite{fengBio}, and is sufficient for almost all practical applications. This is possible because the paper textural patterns are far more distinctive than iris textural patterns. In iris, there exist substantial correlations along the radial structures~\cite{daughmanIRIS2}. The same phenomenon does not exist in paper texture, which explains the higher degrees of freedom in our case. This high level of security makes it possible to simply store the $(r, h, s)$ values on a barcode instead of in a secure database. Alternatively, they may be stored in an RFID chip, and retrieved wirelessly during the verification phrase (e.g., in an e-passport application). 

\begin{figure}
	\centering
	\includegraphics[scale=0.25]{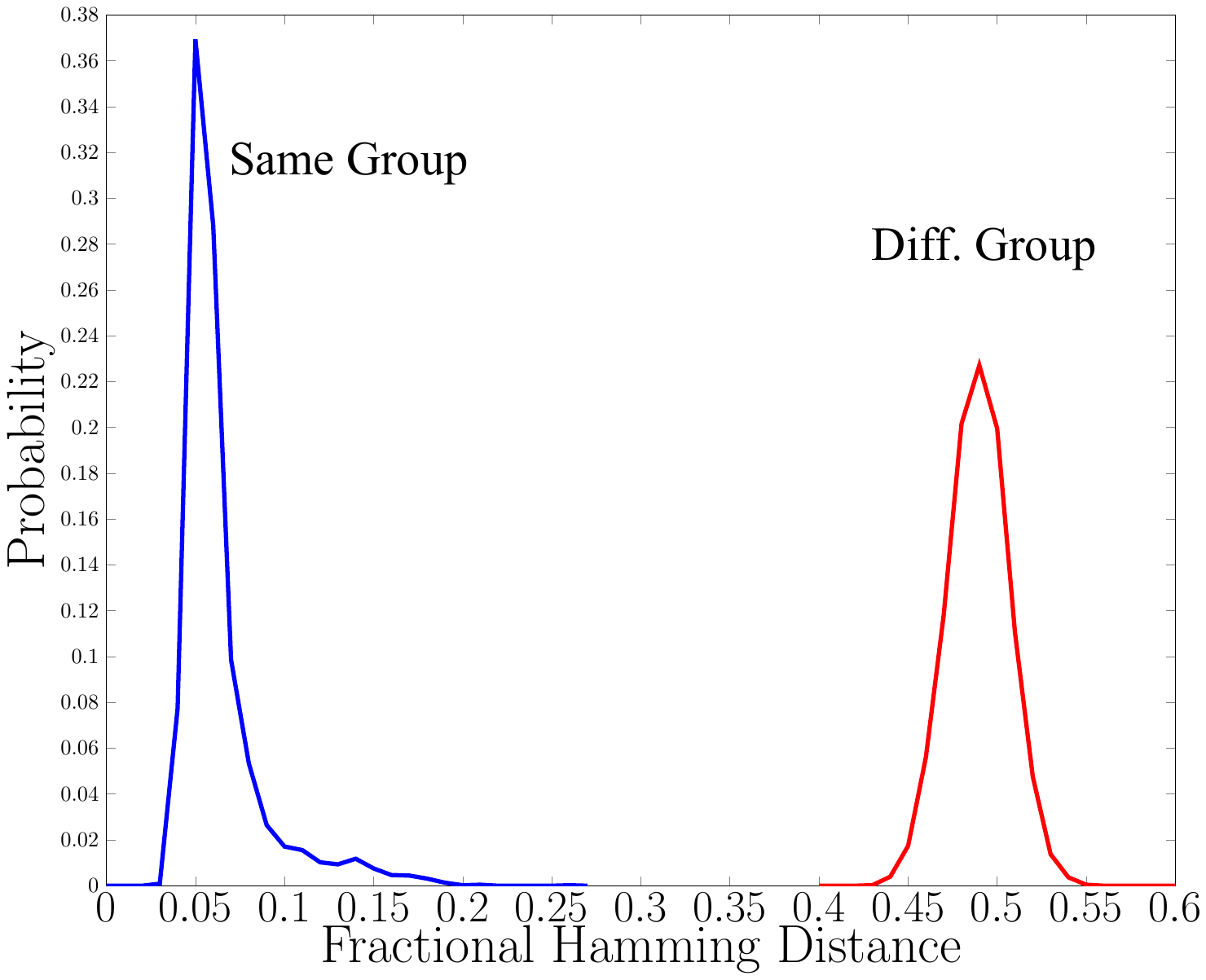}
	\caption{\label{rawHistogram} Histogram of Hamming distances between raw fingerprints without masks.}
\end{figure}

We evaluate the performance of this authentication scheme based on the benchmark database and are able to report perfect error rates: 0\% FRR and 0\% FAR. Note that this performance evaluation is slightly different from the direct comparison between two fingerprints based on their Hamming distance. The authentication is successful, only if the Hadamard-Reed-Solomon code is able to correct the errors (introduced by the XOR between two fingerprints) added to the error correction codeword, and hence recover the same random $k$ (verified again $H(k)$). The authentication protocol can only accommodate raw fingerprints, without masks (see~\cite{fengBio}). Figure~\ref{rawHistogram} shows the histogram of Hamming distance between raw fingerprints without masks. The same-paper and different-paper distributions are well-separated. The error correct code we implemented corrects errors up to 27\%. This is sufficient to correct errors for all same-paper fingerprints, yet not sufficient for different-paper fingerprints. This explains the 0\% FRR and 0\% FAR that we obtain (see Figure~\ref{rawHistogram}).

%

\section{Related work}
\label{background}


In the introduction (Section~\ref{introduction}), we highlight three prominent works in the field (Nature 2005, IEEE S\&P 2009, CCS 2011), which have inspired our work. In this section, we conduct a more comprehensive review of the related work. 



{\bf Special paper.} Some researchers proposed to fingerprint paper by embedding special materials.  
Bauder~\cite{bauder} was the first to propose the idea of a certificate of authenticity (COA), which is a collection of fibers randomly positioned in an object and permanently fixed by using transparent gluing material. Once an end-point of a fiber is exposed to light, the other end is illuminated, and as a whole this creates unique illuminated patterns, which can be captured by a light detector. The main intended application is to use COA for banknotes to ensure authenticity. Kirovski~\cite{fusedCOA} followed up Bauder's work and proposed to combine the captured illuminated patterns with arbitrary text, signed with the private key of the banknote issuer. The signature is then encoded as a barcode and printed on the banknote. 
Chen et al.~\cite{fiberinfused} proposed an improved scanner to achieve automated verification of fiber-based COAs. In a similar work, Bulens et al.~\cite{stronglylinkdata} proposed to embed different material--ultra-violet fibers--into the paper mixture, and use a UV scanner to obtain a unique fingerprint. The authors report that the derived fingerprints have 72-bit entropy.  

{\bf Unmodified Paper and using laser.} One limitation with all the works mentioned above is that they require modifying the paper manufacturing process. Other researchers investigate fingerprinting techniques that can work with ordinary paper without altering the manufacturing process at all. One prominent work along this line of research is due to Buchanan et al.~\cite{paperPUF} published in Nature 2005. The researchers proposed to use a focused laser beam to scan across a sheet of standard white paper and continuously record the reflected intensity from four different angles by using four photodetectors. The laser reflection is random, and is determined by the non-uniform paper surface. Hence, the recorded reflection intensities constitute a unique fingerprint. Beijnum et al.~followed up this research idea in~\cite{sampleCorrelationLaser}. They formulated a criterion for recognition that limits the false acceptance rate (FAR) to 0.1\%. Samul et al.~\cite{laseRecognition} presented a similar idea of shooting a beam of laser onto the surface of the paper. But instead of using photodetectors, they proposed to use a CCD camera to capture the microscopic patterns of speckles.



{\bf Unmodified paper and using light.} Laser-based fingerprinting methods have the limitation that they require special laser equipment. A more cost-effective solution is to use a commonly available light source. Metois et al.~\cite{fiberfingerprint} proposed custom-built equipment called the ``imager'', which consists of a consumer-grade video module and lens, housed along with an embedded lighting apparatus. The imager provides a grayscale snapshot of the naturally occurring inhomogeneities of the paper surface. The snapshot is then processed into a vector of real numbers. The authentication of paper fingerprints is based on computing the correlation coefficient between vectors. The equal error rate (EER) is reported to be about 9\%. 

Clarkson et al.~\cite{clarksonfingerprint} proposed a similar method to fingerprint a paper document by using a commodity scanner instead of a specially built ``imager''. Their work was motivated by the observation that when viewed up close, the surface of a sheet of paper is a tangled mat of wood fibers with a rich three-dimensional texture that is random and hard to reproduce. Utilizing the embedded light emission, the researchers use a commodity scanner to scan a paper sheet in four different orientations. Then a 3-D model is constructed based on these four scans. Furthermore, the 3-D model is compressed into a feature vector through computing Voronoi distributions in the scanned region. The comparison between two feature vectors is based on computing the correlation coefficient.

Pham et al.~\cite{paperfingerprintAlpha} adopted the same approach as Clarkson et al.~\cite{clarksonfingerprint} by using an EPSON 10000XL scanner at 600~dpi to collect 10 scans of the paper surface. In particular, they look at the case when text has been printed over the authentication zone, and propose two methods of pixel inpainting to remove printed text (or marks) from the authentication zone in order to allow ordinary correlation to be performed. Different from the proposed method of Clarkson et al.~\cite{clarksonfingerprint} that compresses the scanned images into a compact feature vector and compares feature vectors based on the correlation coefficient, Pham et al.\ proposed to use alpha-masked imaging matching to compare regions of the two paper surface images. Improvements are demonstrated using the collected data sets in their experiments.

Sharma et al.~\cite{paperspeckle} proposed a different surface-based fingerprinting method. Unlike prior paper fingerprinting techniques~\cite{clarksonfingerprint,paperfingerprintAlpha} that extract fingerprints based on the fiber structure of paper,  their method uses a USB microscope to capture the ``surface speckle pattern'', a random bright and dark region formation at the microscopic level when light falls on the paper surface. The captured patterns are then processed into a vector of digits, which form the unique fingerprint. Fingerprints are compared based on the Euclidean distance between the two vectors.


Beekhof et al.~\cite{beekhof} proposed a fingerprinting method based on measuring random micro-structures of the paper surface. The random micro-structural patterns are captured by using a mobile phone camera with macro lens mounted. The captured image is compressed into a binary feature vector by first hashing the image values into a list of codewords and then running the decoding process through a \emph{reference list decoding} (RLD) technique. Two feature vectors are compared based on the Hamming distance. 


Smith et al.~\cite{smithMicrofiber} proposed another method to capture light reflections from paper surface. Their method involves printing an 8mm box on paper, and then taking a snapshot of it. The alignment can then be done automatically by software based on the printed box, but no details are given in the paper. The authors apply a ``texture hash function'' to generate the fingerprints. Fingerprints are authenticated based on computing correlations of the texture hash strings. 

Haist and Tiziani~\cite{HaistBankNote} proposed a method to fingerprint German banknotes by using a CCD digital camera to take a snapshot of a banknote based on transmissive light. Then, the snapshot is saved as a JPEG image (2.86 KB), which, along with a digital signature, is printed on the banknote as a string of 3250 ASCII characters on an area of 5~cm${}^2$. However, no prototype implementation is reported. The verification is performed by applying the Fourier transform to obtain a feature vector and computing the correlation between the two vectors. Their idea is the closest to ours in terms of using transmissive light. However, our work is substantially different from theirs in several important aspects. First, their work involves testing only three German Deutsche Mark  banknotes while the test data sets used in our work are far more extensive. Second, they do not perform image pre-processing. As a result, the positioning and orientation are done manually rather than automatically by a software algorithm as in our case. Third, they do not carry out image encoding. Consequently, they need to store a JPEG image (2.86 KB), while we only need to store a compact fingerprint (256 byte). Fourth, they do not implement their idea in a prototype system. Hence the feasibility remains uncertain. Most importantly, the Haist-Tiziani paper does not report any error rate performance, or any entropy analysis, and it does not perform extensive robustness tests as we have done.

Renesse~\cite{3dDimensional} proposed a 3-dimensional-structure authentication system (3DAS) to authenticate a standard PVC ID-card that has a 3$\times$3~mm${}^2$ 3DAS-structure in a transparent window. The 3DAS-structure contains spunlaid fibers that are thermally bonded at their cross points. In their experiment setup, two infrared emitting diodes (IREDs) are used as lighting sources to shine on the 3DAS area from two different angles. This creates two shadow images that are then captured by a two-dimensional CCD-array. By alternatively switching both IREDs the required parallel images are produced. Finally, a 20-byte fingerprint is obtained by calculating the centres of gravity of the captured images. However, Renesse's paper does not report error rate performance or perform any entropy analysis. It does not report robustness tests either.

{\bf Summary.} We have presented related paper-fingerprinting techniques proposed in the literature, which have different requirements on paper material, use different types of illuminating sources and scanning equipment, apply different signal processing techniques and obtain fingerprints of different types and features. Our work advances the state-of-the-art in this field by presenting the first practical solution that works with ordinary paper, uses an ordinary lighting source combined with an off-the-shelf camera, takes only 1.3 seconds to produce a compact fingerprint (256 bytes) from one snapshot, achieves an ideal 0\% FFR, 0\% FAR as well as very high entropy (807 bits) in fingerprints, and is demonstrably robust against rotation, crumpling, scribbling, soaking and heating. The near perfect result is attributed to the idea of capturing the paper textural patterns through transmissive light. As detailed in Section~\ref{framework}, using transmissive light reveals richer textural patterns than reflective light and produces more reliable features. This explains our superior result as compared to the earlier surface-based paper fingerprinting methods~\cite{paperPUF,clarksonfingerprint,paperspeckle}.

\section{Conclusion and Future Work}

In this research, we propose to fingerprint a paper sheet based on its texture patterns instead of features on the surface as done by previous work. We show the former contain more distinctive features than the latter with higher decidability in the histogram of Hamming distance distributions. The experiments are set up to use a commodity camera to photograph the texture patterns with a light source shining on the other side of the paper. The rich texture pattens are processed using Gabor wavelets to generate a compact 2048-bit fingerprint code. Based on the collected database, we report zero error rates, and the method is shown to work well with different light sources, and is resistant against various distortions such as crumpling, scribbling, soaking and heating. The extracted fingerprints contain 807 degrees-of-freedom, which is sufficiently high for many practical applications. As an example, some applications (like e-passport) rely on a tamper-resistant RFID chip embedded in the paper document for proving the authenticity of the document (through a challenge-response protocol based on a long-term secret stored in the chip). Our method provides an alternative solution that leverages the natural physical properties of the paper document instead of the tamper resistance of an extra embedded chip.

In this paper, we have focused testing our method on office paper sheets. In future, we plan to extend our study to other types of paper, such as thermal paper, labels and passport pages as long as the light can transmit through. However, based on the thickness of the paper and the difference in the texture materials, we believe it is likely that some changes in the intensity of the light, camera settings, Gabor filter scale and orientation will need to be made. These questions will be addressed in the future work.

\section*{Acknowledgement}

This work is funded by the ERC Starting Grant, No.~306994.

\bibliographystyle{IEEEtran}
\bibliography{PaperFingerprint}

\end{document}